\documentclass[12pt,preprint]{aastex}

\usepackage{amssymb}
\usepackage{amsmath}

\newcommand{\be}{\begin{equation}}
\newcommand{\ee}{\end{equation}}

\shorttitle{Cosmic Scatter and Luminosity Function} 
\shortauthors{Trenti \& Stiavelli}

\begin{document}


\title{Cosmic Variance and Its Effect on the Luminosity Function
Determination in Deep High z Surveys}


\author{M. Trenti} \affil{Space Telescope Science Institute, 3700 San
Martin Drive Baltimore MD 21218 USA} \email{trenti@stsci.edu} \and
\author{M. Stiavelli} \affil{Space Telescope Science Institute, 3700
San Martin Drive Baltimore MD 21218 USA; \\ Department of Physics and
Astronomy, Johns Hopkins University, Baltimore, MD 21218 USA }
\email{mstiavel@stsci.edu}


\begin{abstract}

We study cosmic variance in deep high redshift surveys and its
influence on the determination of the luminosity function for high
redshift galaxies. For several survey geometries relevant for HST and
JWST instruments, we characterize the distribution of the galaxy
number counts. This is obtained by means of analytic estimates via the
two point correlation function in extended Press-Schechter theory as
well as by using synthetic catalogs extracted from N-body cosmological
simulations of structure formation. 
We adopt a simple
luminosity - dark halo mass relation to investigate the environment
effects on the fitting of the luminosity function. We show that in
addition to variations of the normalization of the luminosity
function, a steepening of its slope is also expected in underdense
fields, similarly to what is observed within voids in the local
universe. Therefore, to avoid introducing artificial biases, caution
must be taken when attempting to correct for field underdensity, such
as in the case of HST UDF i-dropout sample, which exhibits a deficit
of bright counts with respect to the average counts in GOODS. A public
version of the cosmic variance calculator based on the two point
correlation function integration is made available on the web.

\end{abstract}


\section{Introduction}

Deep high redshift observations are providing a unique insight into
the early stages of galaxy formation, when the universe was no more
than one billion years old. Large samples of galaxies at $z > 5$ exist
and allow us to study with unprecedented details the early epoch of
galaxy assembly and star formation history (e.g. see
\citealt{ste96,mad96,gia02,bou04,mob05,beck06,oes07}) as well as to
constrain the properties of sources responsible for reionization
(e.g. see \citealt{sti04,yan04,bun06}). Hunting for high-$z$ galaxies
usually follows two complementary approaches. One can either go for
large area surveys, such as GOODS \citep{gia04} and, especially,
COSMOS \citep{sco06}, where the luminosity limit is about $L_{*}$, or
focus on small areas of the sky, such as has been done for the HDF
\citep{will96} and for the UDF \citep{beck06}. In the latter case, the
magnitude limit is $~2.5-3$ mag below $L_*$ (UDF limit) and the aim
is primarily that of probing the faint end of the luminosity
function. While this strategy has a good payoff in terms of galaxy
detections, particularly when the slope of the luminosity function
approaches two near the magnitude limit of the observations, the field
of view of these pencil beam surveys is usually rather small, with the
edge of the order of a few hundred arcseconds. Therefore the number
counts are significantly influenced by cosmic variance (e.g., see
\citealt{som04}). For example, the surface density of i-dropouts at
the GOODS depth found in one of the two HUDF-NICMOS parallel fields
(GO 9803, PI Thompson - identified as HUDFP2 in \citealt{bou06}) is
only one third of the average value from all the $30$ ACS fields in
GOODS \citep{bou06}. Even considering the much larger area of one
GOODS field (about $10' \times 16'$) the expected one sigma
uncertainty in the number counts for Lyman Break galaxies at $z
\approx 6$ is still $\approx 20 \%$ \citep{som04}.

Relatively little effort has been so far devoted at quantifying the
impact of cosmic variance on the determination of properties of high
redshift galaxies. Past studies have generally characterized only the
variance of the number counts of a given cosmic volume
\citep{mo96,col00,new02,som04} or studied the full distribution but
only in numerical simulations appropriate for local galaxies
\citep{sza00}. The impact of cosmic variance on the measurement of the
shape of the luminosity function at high redshift has never been
addressed, despite the fact that it is known that there is an
environmental dependence in the local universe, as $M_*$ in voids is
about one magnitude fainter than in the field \citep{hoy05}. Our goal
is to highlight the problem and to provide methods for addressing it.

In this paper we estimate the variance of the number counts using
extended Press-Schechter analysis and we compute the distribution of
number counts in synthetic surveys generated by using a Monte Carlo
pencil beam tracer in snapshots from cosmological N-body
simulations. Our runs follow the evolution of dark matter particles
only. Galaxy luminosities are linked to the dark matter mass adopting
the parametric models by \citet{val04} and \citet{coo05}, that provide
quite realistic output luminosity function. This numerical approach
allows us to investigate how the best fitting parameters of the
luminosity function depend on the environment probed.

The paper is organized as follows. In Sec~\ref{sec:cv_gen} we define
cosmic variance and we compute it using the extended Press-Schechter
formalism. In sec.~\ref{sec:sint} we describe the numerical framework
that we adopted to investigate cosmic variance. In sec.~\ref{sec:cv}
we apply the Monte Carlo code to characterize the number counts for a
variety of high-z surveys strategies, while in sec.~\ref{sec:lf} we
investigate the impact of number counts variance on the determination
of the luminosity function, focusing in particular on the faint end
slope of i-dropout galaxies in the UDF main and parallel fields. We
give our conclusions in sec.~\ref{sec:conc}.

\section{Large scale structure and uncertainties in galaxy number counts}\label{sec:cv_gen}

The number counts of galaxies in a survey are affected by a
combination of discrete sampling, observational incompleteness and
large scale structure. Given a probability distribution $p(N)$ for the
number counts with mean $\langle N \rangle$ and variance $\langle N^2
\rangle$, we define the total fractional error of the counts $N$ as:
\begin{equation}\label{eq:frac_err}
v_r = \frac{\sqrt{\langle N^2 \rangle - \langle N \rangle ^2}}{\langle N \rangle}.
\end{equation}
The uncertainties in excess to Poisson shot noise are usually
quantified in terms of a relative ``cosmic variance'' (e.g. see
\citealt{new02,som04}) defined as:
\begin{equation}\label{eq:cv}
\sigma_{v}^2 \equiv \frac{\langle N^2 \rangle - \langle N \rangle^2}{\langle
N \rangle^2} - \frac{1}{\langle N \rangle}.
\end{equation}

To evaluate the relative cosmic variance $\sigma_v^2$ of a given
sample, two main theoretical approaches are possible: (i) estimation
based on the two point correlation function $\xi(r)$ of the sample or
(ii) direct measurement using mock catalogs from cosmological simulations
of structure formation.

In the first case $\sigma_v^2$ is derived from $\xi(r)$ as follows
(e.g. see \citealt{peebles}):
\begin{equation} \label{eq:sigma_from_xi}
\sigma_{v}^2 = \frac{\int_V \int_V d^3x_1 d^3x_2 \xi(|\vec{x_1}-\vec{x_2}|)}{\int_V \int_V d^3x_1 d^3x_2},
\end{equation}
where the integration is carried out over the volume $V$ observed by
the survey. The two point correlation function $\xi(r)$ can either be
derived from a cosmological model for the growth of density perturbations
(e.g. see \citealt{new02}), or from a simpler analytical model, such
as a power law, with free parameters fixed by observational data
\citep[e.g. see][]{som04}. This method has the advantage of requiring
little computational resources, limited to the evaluation of the
multidimensional integral in Eq.~\ref{eq:sigma_from_xi}, while being
able to handle an arbitrarily large survey volume of any geometry. The
main limitation is that only the second moment of the counts
probability distribution $p(N)$ can be obtained. This method also
relies on an input two point correlation function which is typically
evaluated only considering linear evolution of density perturbations
(see however \citealt{pea96} for an analytical model of the non linear
evolution of the power spectrum). In addition this framework assumes
that all the survey volume is at a given redshift, which may not be
appropriate for high redshift ($z>5$) Lyman Break Galaxies
observations, where the pencil beam extends for about $\Delta z
\approx 1$, an interval over which there is evidence of evolution of
the galaxy luminosity function \citep[e.g., see][]{bou06}.

A direct measurement of $p(N)$ from N-body simulations bypasses all these
limitations but requires significant computational resources. In
particular the cosmic volume simulated must be much larger than the
survey volume, which in practice sets a limit on the survey area that
can be adequately modeled. 

To summarize, if one is interested primarily in computing a total
error budget on the number counts of a survey, the estimate of
$\sigma_v^2$ through integration of $\xi(r)$ may be sufficient and
requires relatively little effort. However in this paper we combine
both methods, emphasizing especially the construction of mock catalogs
from N-body simulations, as this is required to address the influence
of cosmic variance on the uncertainty in the shape of the luminosity
function.

\subsection{Cosmic variance in Extended Press Schechter Theory}\label{sec:cvEPS}

Following \citet{new02} we estimate the cosmic variance in linear
theory evaluating the integral of Eq.~\ref{eq:cv} using $\xi(r)$
derived from the transfer function of \citet{eis99}. For a given
redshift $z$, the halo-dark matter bias $b(M_h)$ as a function of the
halo mass ($M_h$) is evaluated using the \citet{she99} formalism. The
total average bias of the sample is then computed by averaging
$b(M_h)$ over the \citet{she99} mass function down to a mass limit set
by matching the desired comoving number density of halos.

These are the basic steps to evaluate the influence of cosmic variance
in the error budget of the number counts using linear theory.

\begin{itemize}

\item \emph{Define the survey volume and its average redshift
$z_{av}$}. For example, for Lyman Break Galaxies dropouts samples,
this is set by the combination of the field of view angular size and
of the redshift window for the dropout selection.

\item \emph{Choose the intrinsic number of objects in the survey}. For
example, this can be done either starting from a specific luminosity
function or estimating the number of expected objects from the actual
number of observed objects divided by their completeness ratio.

\item \emph{Estimate the average incompleteness}. Incompleteness will be
close to 0 for selections much brighter than the magnitude limit of
the survey but can be in the range $0.3-0.5$ when pushing detections
up to the limit of the data \citep{oes07}. A precise estimate
generally requires object recovery Monte Carlo simulations.

\item \emph{Adopt a value for the average target - halo filling
factor}. This is in general smaller than 1, as a specific class of
objects may be visible only for a limited period of time. For example
in the case of Lyman Break Galaxies, the duty cycle may be as low as
$0.25$ \citep[e.g. see][]{ver07}. 

\end{itemize}

The input information above is then used to estimate the total
fractional uncertainty on the number counts as follows. 

\begin{itemize}

\item \emph{Compute the minimum halo mass $M_{min}$ required to obtain
the number density of halos hosting the survey population}. Combining
halo filling factor and intrinsic number of objects in the survey
(given the survey volume) we compute the minimum halo mass in the
\citet{she99} model required to match the input number density.

\item \emph{Compute the average bias of the sample}. We calculate the
average bias of the sample using Press-Schechter model \citep{ps74}.

\item \emph{Integrate the dark matter $\xi(r)$ over the survey
volume}. The dark matter two point correlation function is integrated
over the pencil beam geometry of the volume
(cf. Eq.~\ref{eq:sigma_from_xi}, see also \citealt{new02}) to obtain
the dark matter cosmic variance $\sigma^2_{DM}$.

\item \emph{Multiply $\sigma_{DM}$ by the average galaxy bias}
to obtain the cosmic variance of the sample: $\sigma_v^2= b^2
\sigma^2_{DM}$.

\item \emph{Take into account Poisson noise for the number of observed
objects}.  The total error budget is given by combining the
contribution from cosmic variance, which is an intrinsic property of
the underlying galaxy population, with the observational uncertainty
related to the actual number of observed objects $N_{oss}$. Therefore
the total fractional error (that is the one sigma uncertainty) is:
\begin{equation}
v_r = \sqrt{\sigma_v^2+1/N_{oss}}.
\end{equation}

\end{itemize}

In Sec.~\ref{sec:cv} we use this method to estimate the number
counts uncertainty for typical high redshift surveys.

\section{Synthetic surveys}\label{sec:sint}

In this section we present the numerical framework based on
cosmological simulations that we developed to address the influence of
cosmic variance on high redshift observations.

\subsection{N-body simulations}

The numerical simulations have been carried out using the public
version of the PM-Tree code Gadget-2 \citep{spr05}. We adopt a
cosmology based on the third year WMAP data \citep{WMAP3}:
$\Omega_{\Lambda}=0.74$, $\Omega_{m}=0.26$, $H_0=70~$km/s/Mpc and
$\sigma_8=0.75$ (or $\sigma_8=0.9$, see later), where $\Omega_m$ is
the total matter density in units of the critical density ($\rho_{c}=
3H_0^2/(8 \pi G)$) with $H_0$ being the Hubble constant (parameterized
as $H_0 = 100 h$ km/s/Mpc) and $G$ the Newton's gravitational constant
\citep{peebles}. $\Omega_{\Lambda}$ is the dark energy density and
$\sigma_8$ is the root mean squared mass fluctuation in a sphere of
radius $8$~Mpc/h extrapolated to $z=0$ using linear theory.  The
initial conditions have been generated with a code based on the Grafic
algorithm \citep{bert01} using a $\Lambda CDM$ transfer function
computed via the fit by \citet{eis99} with spectral index $n_s=1$.

A summary of our N-body runs is presented in Table~\ref{tab:sim}. We
resort to various box sizes, optimized for different survey
geometries. To characterize small area surveys such as a single ACS
field, we use a box of edge $100$~Mpc/h simulated with $512^3$
particles. This choice gives us a volume about $73$ times larger than
the effective volume probed by one ACS field for V-dropouts ($\approx
5.7 \times 5.7 \times 420$ (Mpc/h)$^3$) and about $86$ times larger for
i-dropouts ($\approx 6 \times 6 \times 320$ (Mpc/h)$^3$). The single
particle mass is $5 \cdot 10^8 M_{\sun}/h$ and guarantees that the
host halos that we consider ($M_{halo} \gtrsim 2 \cdot 10^{10}
M_{\sun}/h$) are resolved with at least $40$ particles in the deepest
survey that we simulate. To investigate the cosmic variance in larger
area, less deep surveys, such as GOODS, we use instead a larger box,
of size $160$ Mpc/h, also simulated with $512^3$ particles. The total
volume of the simulation is about 26 times larger than a single
i-dropouts GOODS field ($\approx 17.6 \times 28.1 \times 320$
(Mpc/h)$^3$). The single particle mass for this run is $2 \cdot 10^9
M_{\sun}/h$. These simulations have $\sigma_8=0.75$. In addition, we
consider a higher resolution simulation with more than twice the
number of particles of our basic runs ($N=680^3 \approx 3.1 \cdot
10^8$) and with a box of edge $128 Mpc/h$ that starts with
$\sigma_8=0.9$. This simulation has a single particle mass of $4.4
\cdot 10^8 M_{\sun}/h$.
 
Dark matter halos are identified in the simulations snapshots (saved
every $\Delta z =0.125$ in the redshift interval $10-4.5$) using the
HOP halo finder \citep{eis98}, with the following parameters: the
local density around each particle is constructed using a 16 particles
smoothing kernel, while for the regrouping algorithm we use
$\delta_{peak} = 200$, $\delta_{saddle} = 170$, $\delta_{outer} = 100$
and a minimum group size of $40$ particles.  The halo mass
distribution in the snapshots is well described (with displacements
within $\approx 25\%$) by a \citet{she99} mass function.
One limitation that has to be taken into account when estimating the
cosmic variance from N-body simulations is the cosmic variance of the
complete simulation volume. For example, in our $512^3$ particles,
$100$ Mpc/h edge box run there are 2620 dark matter halos at $z=6$
with more than 100 particles and the differences in this number from
run to run are larger than the nominal Poisson variance. We checked
this effect by carrying out a control run with the same initial
conditions but a different seed for the random number generator
obtaining a difference of $\approx 6\%$. The use of a larger box and
$680^3$ particles in our highest resolution run is expected to reduce
run to run variations, but we cautiously assume a $10\%$ relative
uncertainty on the value of the cosmic variance that we derive through
this paper.  

In addition, in the $680^3$ simulation we also save snapshots at
intervals $\Delta z =0.5 $ from $z=10$ to $z=15$ in order to provide a
preliminary characterization of cosmic variance in future JWST NIRCam
surveys.

\subsection{Pencil beam tracing}

Our pencil beam tracer is similar to the one developed by
\citet{kit06}, but it is optimized for high $z$ surveys, allowing us
in particular to take advantage of the quasi-constant angular distance
versus redshift relation. We trace through the simulation box a
parallelepiped where the base is a parallelogram, whose size is given
by the field of view of the survey in comoving units, and the depth is
the comoving depth associated with the redshift uncertainty of the
selection window for Lyman Break Galaxies which we are interested
in. This choice means that we are neglecting the variation of angular
distance versus redshift in the redshift interval of the selection
window considered. For example, the comoving edge of the ACS field of
view for V-dropouts is $5.5$ Mpc/h at $z=4.6$ and $5.9$ Mpc/h at
$z=5.7$, and we approximate it with $5.7$ Mpc/h. The pencil beam is
traced through different snapshots as we swipe through its depth, and,
consistently with our angular size approximation, we assume an average
value for $\Delta z$ (the redshift difference between two snapshots)
expressed in comoving distance. The choice to save snapshots at
$\Delta z =0.125$ implies that a single beam passes through several
different snapshots so that the evolution in the halo mass function is
well captured. $\Delta z = 0.125$ is equivalent to about $40$ Mpc/h at
$z=6$ and the evolution in the number density of halos at the same
mass scale between two adjacent snapshots is of the order $10 -15
\%$. The beam starts at a random position within the simulation volume
and then proceeds through the cube with periodic boundary conditions,
angled using the following choices for the two direction angles
$\theta_1$ and $\theta_2$:
$$
\tan{\theta_1} = 0.3
$$
and
$$
\tan{\theta_2} = 0.5.
$$ 
These values have been selected to guarantee no superposition and
adequate spacing for a typical HST-ACS Lyman Break dropouts beam as it
wraps around the simulation box due to the periodic initial
conditions. The linear correlation between the number counts of two
nearby segments of the beam, estimated using the two point correlation
function (see Sec.~\ref{sec:cvEPS}) is in fact $\lesssim 0.02$.

Finally all the dark matter halos within the beam are flagged and
saved for subsequent processing by the Halo Occupation Distribution
part of the Monte Carlo code. A complete description of the properties
of the pencil beam surveys simulated in this paper is given in
Tab.~\ref{tab:fields}.


\subsection{Halo Occupation Distribution}

Dark matter halos within a simulated field of view are populated
with galaxies accordingly to a simple Halo Occupation Distribution
(HOD) model. We assume an average occupation number \citep{wec01}:
\begin{equation}
\langle N_{gal} \rangle = \theta(M-M_{min})(1 + (M/M_1)^{\beta}),
\end{equation}
where $\theta(x)$ is the Heaviside step function, $M_{min}$ is a
minimum halo mass threshold, $M_1$ is the typical scale where multiple
galaxies are present within the same halo and $\beta>0$ is of the
order unity. If $M>M_{min}$, one galaxy is placed at the center of the
halo and then the number of companions is extracted from a Poisson
distribution with mean $(M/M_1)^{\beta}$. Of these galaxies a fraction
$f_{ON}$ is finally identified as Lyman Break Galaxies (this is to
take into account possible observational incompleteness and/or a duty
cycle where LBGs are on only for a fraction of their lifetime).

\begin{figure} 
  \plotone{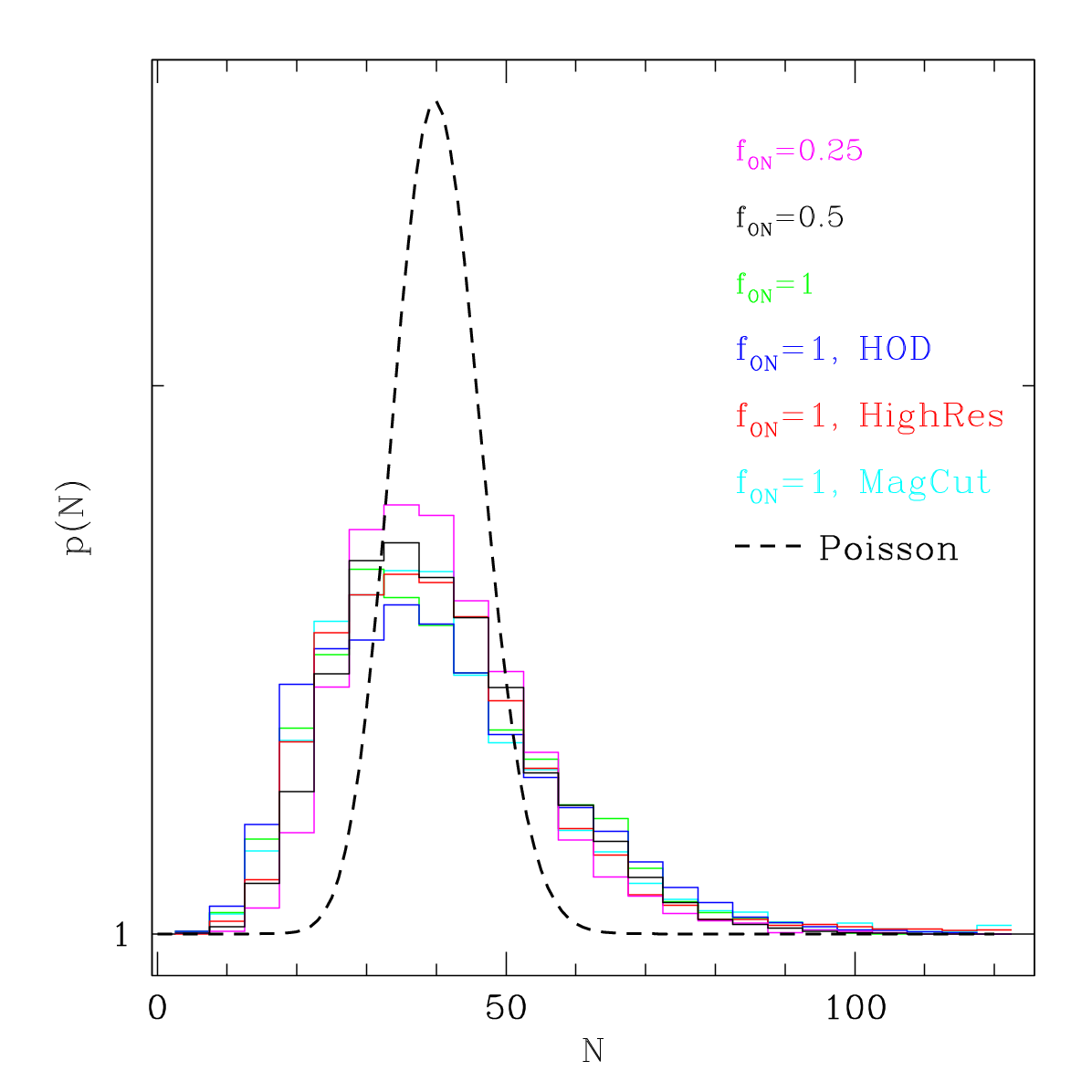} \caption{Probability distribution of the
  i-dropouts number counts for an ACS field with 40.3 average counts,
  using different Halo Occupation Models and different simulations
  boxes: $N=512^3$ and $N=680^3$ (HighRes). The cosmic variance for
  these runs is quantified in table~\ref{tab:uncert}. For reference
  the Poisson distribution for 40.3 counts on average is plotted with
  a dashed line. }\label{fig:idrop_models}
\end{figure}

We show in Fig.~\ref{fig:idrop_models} the distribution of the number
counts for i-dropouts in one ACS field at the UDF depth.  We consider:
(i) $f_{ON}=1$, $M_1 \to +\infty$ (that is one galaxy per halo); (ii)
$f_{ON}=0.5$ and $f_{ON}=0.25$, $M_1 \to +\infty$ (that is one galaxy
per halo with either 0.25 or 0.5 detection probability); (iii)
$f_{ON}=1$, $M_1 = 5\cdot 10^{11} M_{\sun}/h$ and $\beta=1$. In
addition we also investigate the effect of changing the box size, the
resolution and the $\sigma_8$ value of the N-body simulations using
both our $N=512^3$ and our $N=680^3$ runs. The value of $M_{min}$ is
kept as free parameter and it is adjusted to have the same average
number of counts in all three cases. The required variations of
$M_{min}$ are limited within a factor two. The distribution of the
number counts probability is very similar in the four cases (see
table~\ref{tab:uncert} for $v_r$ values) and shows that the
characterization of the cosmic variance versus the average number
counts is solid with respect to the details of the modeling. This is
essentially due to the fact that by changing $M_{min}$ by a factor of
2, the average bias of the sample varies only by less than
$15\%$. This is reassuring as there are large theoretical
uncertainties on modeling high redshift galaxy formation.

\subsection{Field of view geometry}

\begin{figure} 
  \plotone{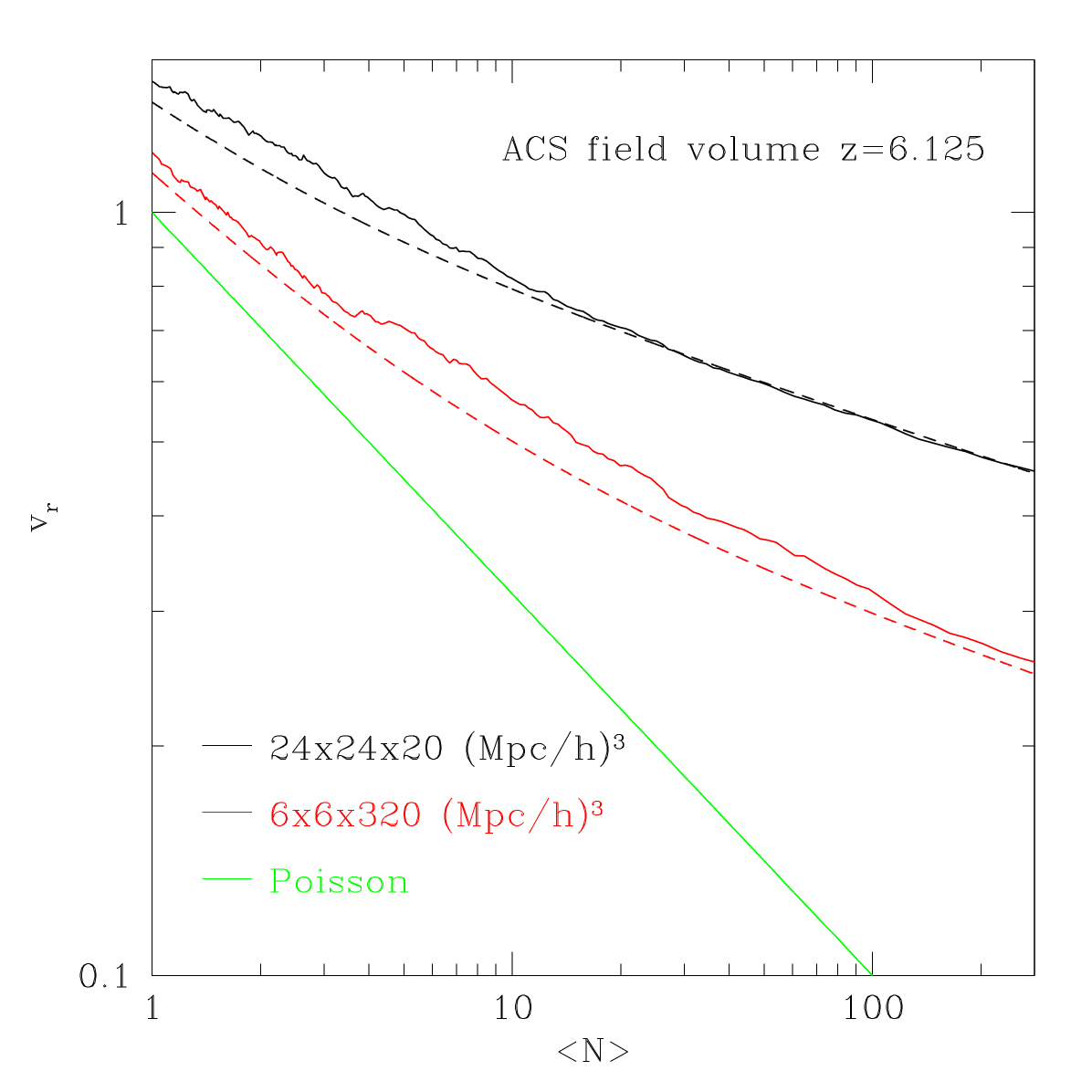} \caption{Total fractional error
  $v_r$ of the number counts of dark matter halos in a cosmic volume
  of $1.152 \cdot 10^{4} (Mpc/h)^3$ at $z=6.125$ as a function of the
  average number counts for different shapes of the volume. The solid
  line is the measurement from our numerical simulations while the dashed
  line is the prediction using the analytic model described in
  Sec.~\ref{sec:cvEPS}.  The quasi-cubic volume has the highest
  standard deviation, especially at large number counts, where the
  Poisson noise (green line) has a negligible contribution to the
  total fractional error (see also
  Fig.~\ref{fig:sigma_poi}).}\label{fig:shape}
\end{figure}

The importance of modeling correctly the geometry of the field of view
to quantify the cosmic variance is shown by Fig.~\ref{fig:shape}. Here
we focus on a single snapshot at $z=6.125$ and we measure the total
fractional error $v_r$ of number counts in a volume of $11520
(Mpc/h)^3$ with different shapes. From Fig.~\ref{fig:shape}, $v_r$ is
largest in the quasi-cubical volume and smallest for the pencil
beam. Therefore cosmic variance computed only from the total volume of
the survey, assumed to be a sphere as in \citet{som04}, is
overestimated. This is because a narrow and long pencil beam probes
many different environments, while a cubic volume may sit for a
significant fraction of its volume either on under-dense or on
over-dense regions. In all cases shown in Fig.~\ref{fig:idrop_models},
the total error exceeds Poisson noise and the contribution from cosmic
variance is dominant especially when the average number of counts in
the volume is much larger than one. Fig.~\ref{fig:shape} also shows an
excellent agreement between the measurement from cosmological simulations
and the analytical estimate using the two point correlation function.

\subsection{Luminosity-Mass relation}\label{sec:lm}

Our numerical simulations follow the evolution of dark matter only,
therefore we have to assume a mass-luminosity relation to investigate
the influence of the cosmic variance on the determination of the
luminosity function. For this we extend to higher redshift the fitting
formulas used by \citet{val04} and by \citet{coo05} adopting:
\begin{equation}\label{eq:lm}
L(M)= L_{0} \frac{(M/M_s)^{a}}{[q+(M/M_s)^{cd}]^{1/d}},
\end{equation}
and assuming $M_s=10^{11}M_{\sun}$, $a=4.0$, $q=0.57$, $c=3.57$ and
$d=0.23$. $L_0$ is conventionally set to $1$. As the main goal of this
paper is not to provide a detailed modeling of the mass - luminosity
relation, but rather to investigate the dependence of the luminosity
function parameters on the environment and on the fitting method
adopted, the simple modeling of eq.~\ref{eq:lm} appears adequate.

For a fully consistent representation of the observational selection
process for Lyman Break Galaxies, it would be necessary to apply a
luminosity cut in the apparent and not in the absolute magnitude. The
apparent luminosity-distance relation does in fact evolve quite
significantly over a redshift range $\Delta z \approx 1$ (e.g. see
\citealt{bou06} Fig. 7). This effect leads to a reduction of the
effective volume of the survey and thus to an increase of the cosmic
variance contribution to the counts uncertainty. This is because
dropouts tend to be preferentially selected in the low redshift corner
of the selection window, even if luminosity evolution with redshift
will partially offset this trend as galaxies at higher $z$ appear to
have a higher $L/M$ ratio (see \citealt{cooray05}). In our standard
model for number counts uncertainty from numerical simulations we
apply a cutoff in mass (that is in absolute magnitude), which gives us
a framework consistent with the estimates from the two point
correlation function. To evaluate the error introduced by this
assumption we have implemented a cutoff in observed magnitude for an
ACS i-dropout sample with 40 detections on average constructed from
our cosmological simulation at high resolution and using the relation
between observed $L_*$ and redshift $z$ as plotted in Fig.~7 of
\citet{bou06}. The resulting number counts distribution is plotted in
Fig.~\ref{fig:idrop_models}. The total fractional uncertainty in the
number counts is only marginally higher ($0.42$ vs. $0.40$) than in
the case where a cutoff in absolute magnitude is applied. 

\section{Cosmic Variance}\label{sec:cv}

\subsection{HST surveys}

\begin{figure} 
  \plotone{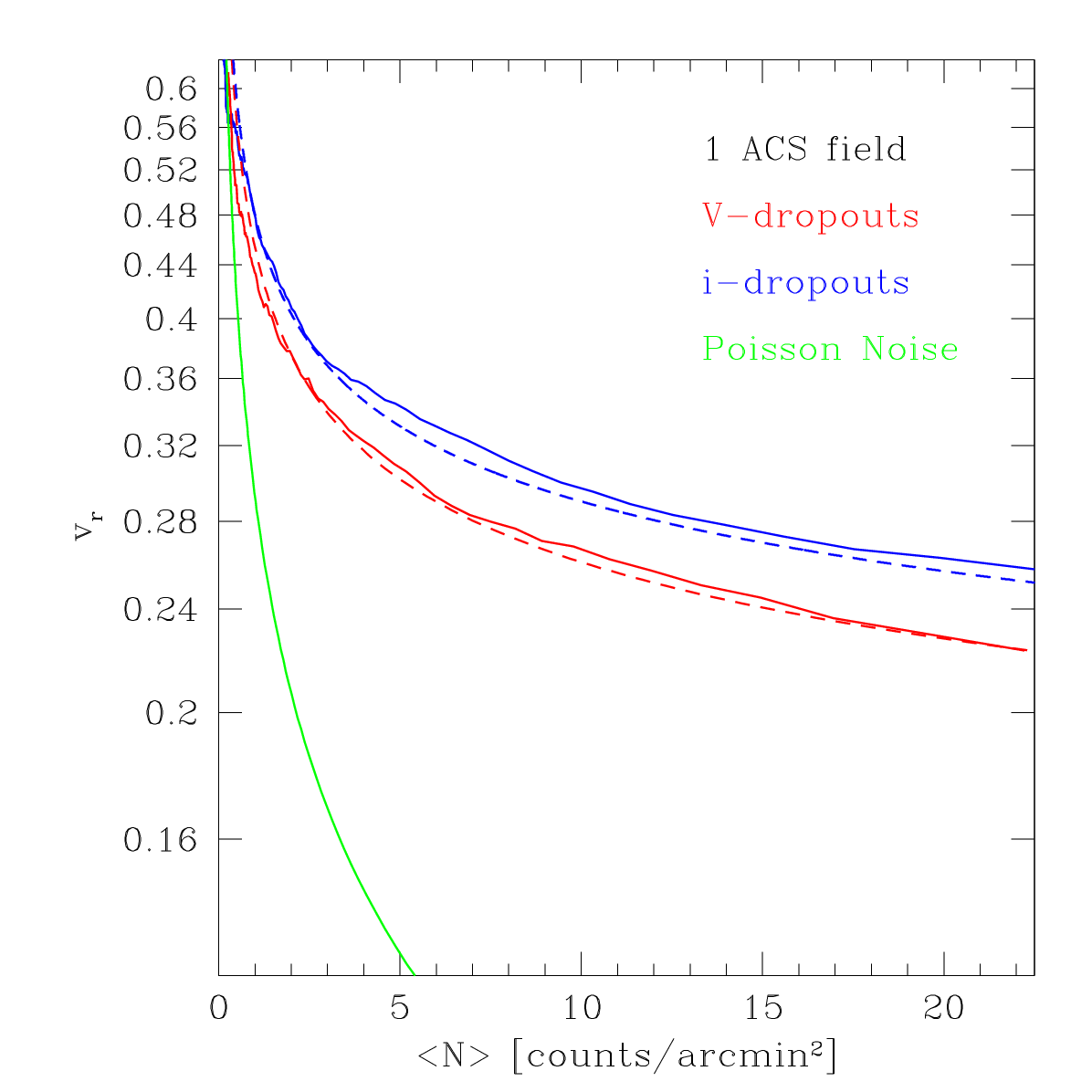} \caption{Total fractional uncertainty
  $v_r$ of the number counts estimated using model (dotted) and
  simulations (solid) for V (red) and i (blue) dropouts for
  one ACS field (in units of the average number of counts) versus the
  average number of counts per $arcmin^2$. The Poisson noise
  associated to the counts is plotted for reference in
  green.}\label{fig:Vidrop}
\end{figure}

\begin{figure} 
  \plotone{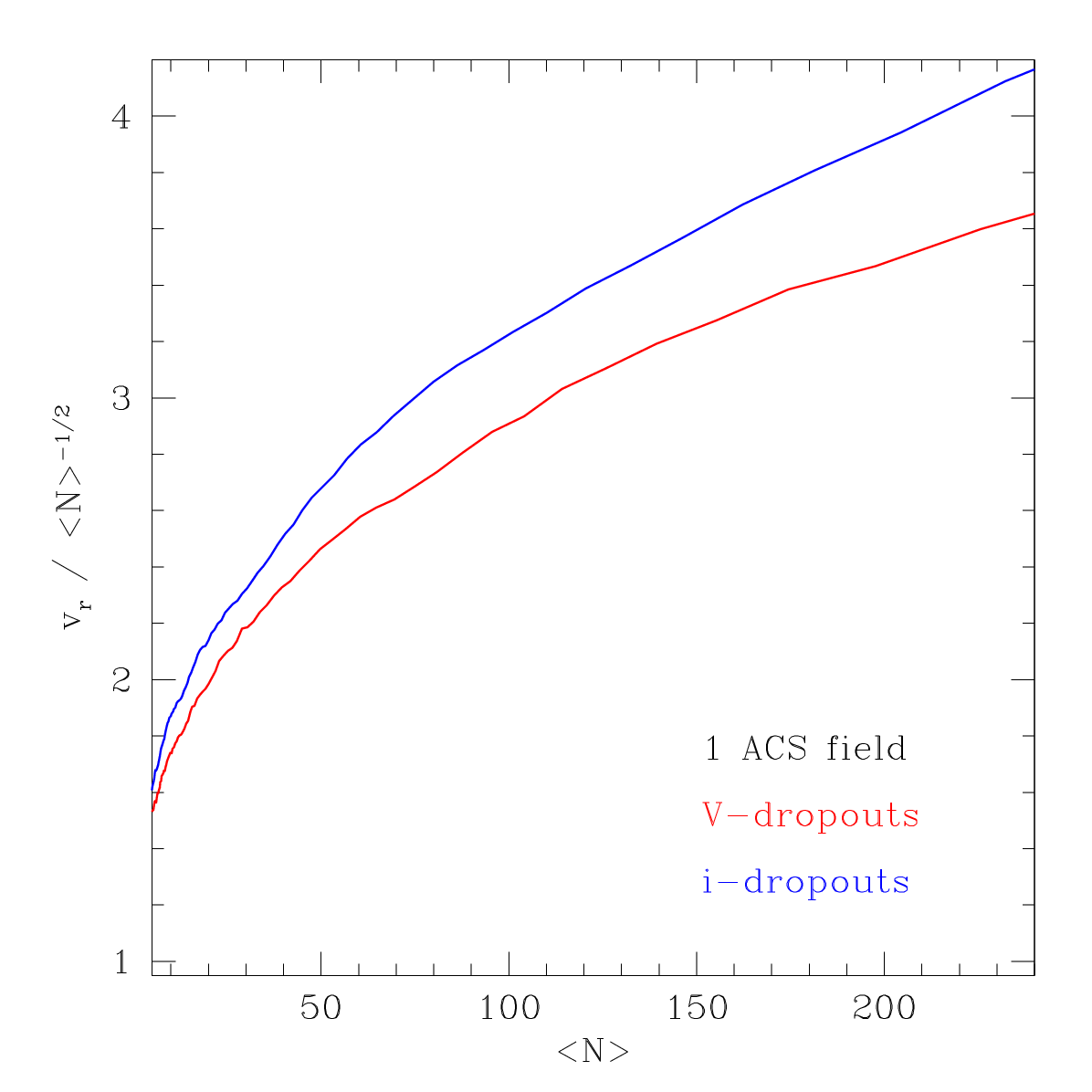} \caption{Total fractional error ($v_r$) in
  the number counts for V and i-dropouts in one ACS field in units of
  the Poisson noise ($1/\sqrt{\langle N \rangle}$) versus the average
  number of counts ($\langle N \rangle$) per field. As $v_r$ is much
  bigger than the corresponding Poisson noise, cosmic variance is the
  dominant source of uncertainty in the number counts.}\label{fig:sigma_poi}
\end{figure}

\begin{figure} 
  \plotone{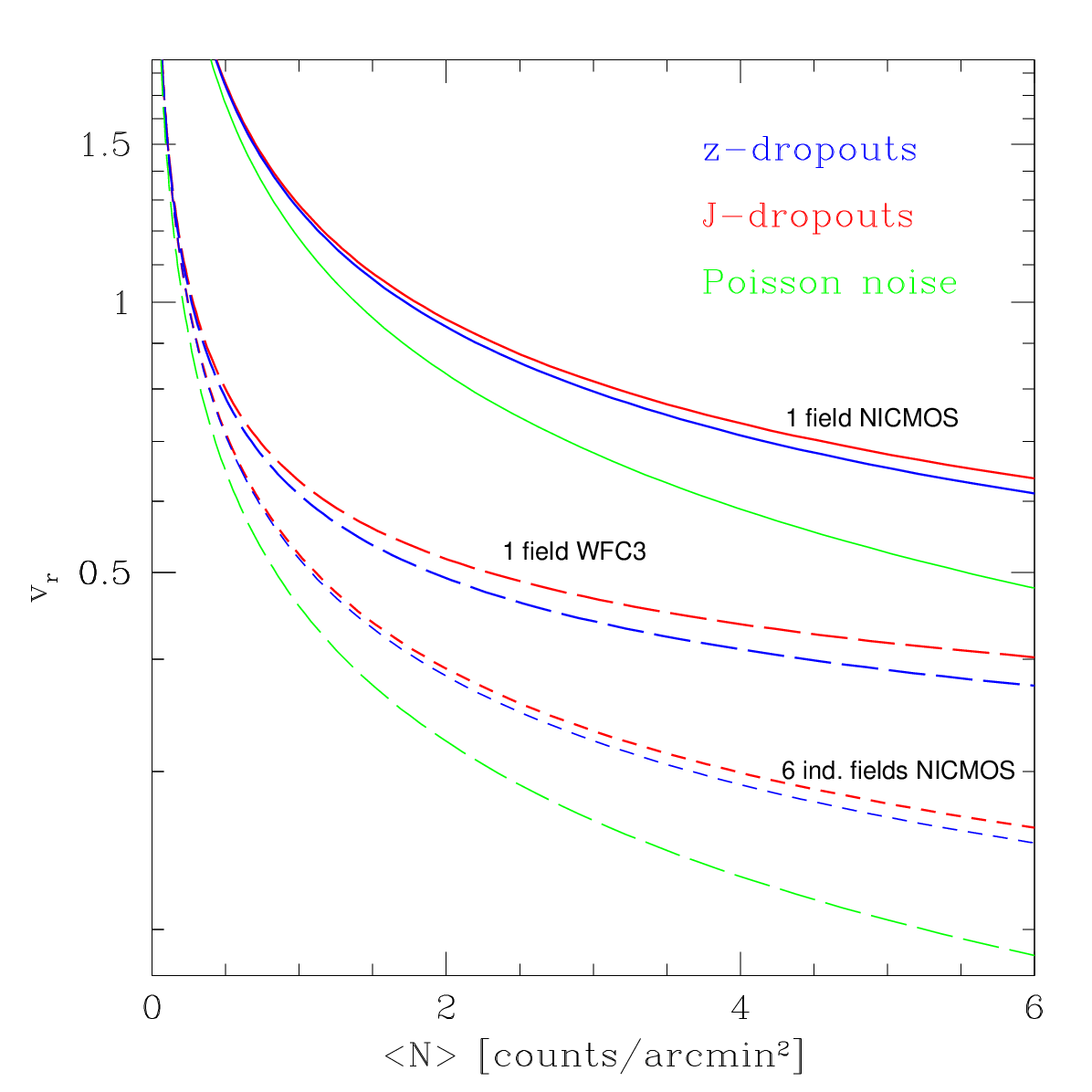} \caption{Like in Fig.~\ref{fig:sigma_poi} (all
  curves from analytical modeling) but for z and J-dropouts
  considering different fields of view: (i) NICMOS Camera 3 (solid),
  (ii) 6 independent NICMOS Camera 3 fields summed (short dashed),
  (iii) one WFC3 field (long dashed). Poisson noise for a single
  NICMOS Camera 3 and WFC3 field is shown in green.}\label{fig:zJdrop}
\end{figure}

The results from our simulated distribution of number counts for
different high redshift surveys are reported in
Figs.~\ref{fig:Vidrop}-\ref{fig:zJdrop}. For a given survey geometry
we plot the fractional number counts uncertainty $v_r$ as a function
of their average value. This facilitates the application to surveys at
different depths. Given that the details of the Halo Occupation Model
have only a modest influence on the value of $v_r$ (see
Fig.~\ref{fig:idrop_models}), we resort in this paper to a reference
model of one galaxy per halo, with $f_{ON}=1$ (i.e. unit probability
of detection) unless otherwise noted. In addition, we highlight the
expected total fractional uncertainty on the number counts due to
Poisson noise only.

At the typical number counts for a field at the UDF depth the total
fractional uncertainty is $v_r \approx 25\%$ for V-dropouts (assuming
100 detections per field, see \citealt{oes07}) and $v_r \approx 35\%$
for i-dropouts (assuming 50 detections per field, see
\citealt{beck06}). This is much smaller than the Poisson noise
associated to the average realized number counts. Therefore cosmic
variance is the dominant source of uncertainty for UDF-like deep
fields. At lower number counts per field of view, that is when the
survey is shallower, the Poisson contribution to the total fractional
uncertainty $v_r$ increases (see Fig.~\ref{fig:sigma_poi}) until it
becomes dominant at the limit of zero average counts. The agreement
between $v_r$ estimated using the two point correlation function and
that measured in the numerical simulations is very good. The framework
is also consistent with the observed clustering of high-redshift
galaxies, such as measured by \citet{ove06}: our standard recipe for
populating dark matter halos with galaxies gives an average
galaxy-dark matter bias $b = 5.0$ for an ACS i-dropout sample with 60
detections on average, a value within the one sigma error bar in the
measurement by \citet{ove06}. Note however that the large
observational uncertainties do not allow a more detailed quantitative
comparison.

From the full probability distribution on the number counts, derived
from our simulations, we can also evaluate the likelihood of finding a
factor 2 overdensity in galaxies at $z=5.9 \pm 0.2$ in the UDF field
as reported by \citet{mal05}. If we assume that the expected number of
galaxies in that redshift interval is 7.5, the fractional uncertainty
in the counts turns out to be $ v_r \approx 67\%$ and the probability
of $15$ or more realized counts in that redshift interval is greater
than $10\%$ (this has been measured taking advantage of the full
probability distribution obtained through the simulations). Therefore
the measured fluctuation, while lying outside the standard deviation,
is not exceptional. Given that a i-dropouts selection window contains
about 5 intervals of width $\Delta z =0.2$, one overdensity such as
that observed by \citet{mal05} will be present in a significant number
of random pointings.

\begin{figure} 
  \plotone{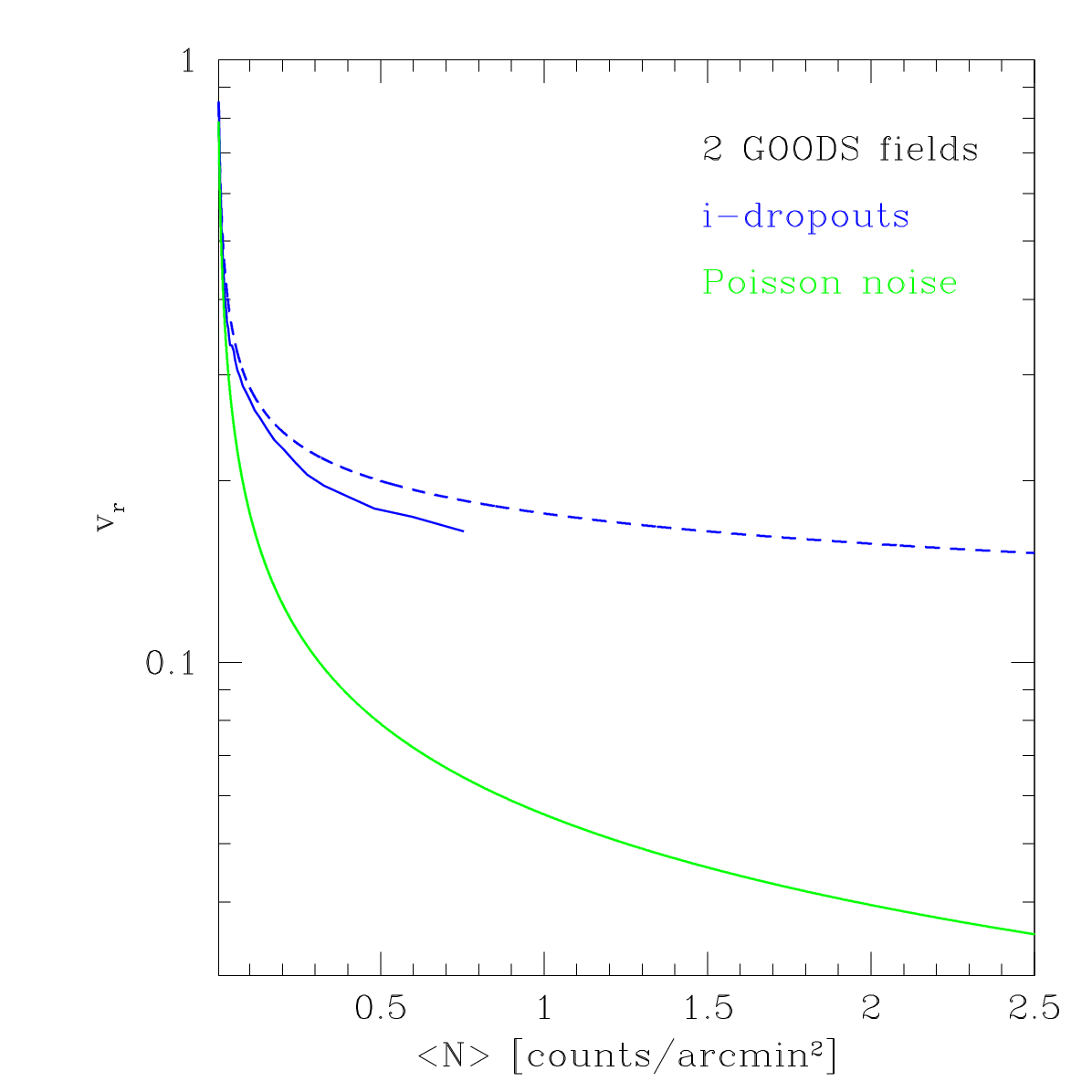} \caption{Total fractional error for i-dropouts
  in the two GOODS fields (solid line: simulations; dashed line:
  model) versus the average number of counts per arcmin$^2$. Poisson
  noise is plotted for reference in green.}\label{fig:idropGOODS}
\end{figure}

To characterize the number counts uncertainty in the larger area GOODS
survey we resort to the simulation with edge $160 Mpc/h$, whose
results are reported in Fig.~\ref{fig:idropGOODS} for the combination
of the two North and South fields. $v_r$ is consistent with the
estimate using the two point correlation function within a relative
difference of $10 \%$ at most. 
Interestingly, the i-dropouts cosmic variance in one GOODS field is
not too different from the one for a smaller but deeper UDF
field. This is due to two effects: (i) the increased sensitivity of
the UDF enables to detect fainter Lyman Break galaxies and thus to
probe the distribution of smaller mass halos that are progressively
less clustered, (ii) there is a significant correlation between the
counts in adjacent fields. 

\begin{figure} 
  \plotone{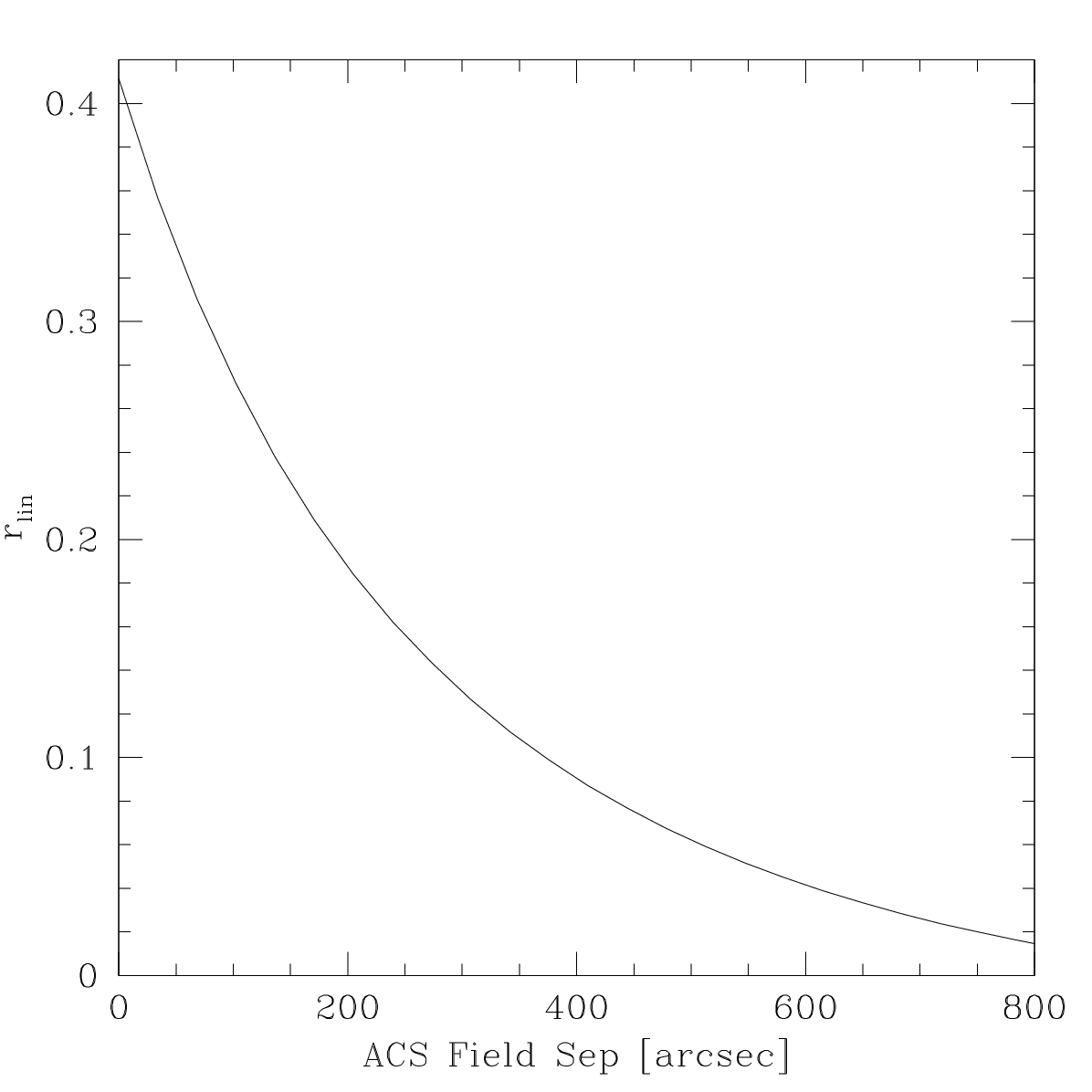} \caption{Linear correlation coefficient
  $r_{lin}$ for the number counts of i-dropouts in two nearby ACS
  fields with an average of 50 detections per field as function of
  their separation.}\label{fig:corr}
\end{figure}

To better quantify this latter effect, we plot in Fig.~\ref{fig:corr}
the linear correlation coefficient between the i-dropouts counts in
two nearby ACS fields:
\begin{equation}
r_{lin} = \frac{\sigma_{12}}{\sqrt{\sigma_{11} \sigma_{22}}},
\end{equation}
where
\begin{equation}
\sigma_{ab} = \langle (a b - \langle ab \rangle)^2 \rangle.
\end{equation}
The linear correlation has been computed using the analytical model
from the two point correlation function. 
Two adjacent i-dropouts fields at the UDF depth (50 counts on average)
have a linear correlation in the number counts of about $0.4$. The
correlation decreases rapidly as the separation increases and fields
separated by more than $500$ arcsec have $r_{lin}<0.05$. Therefore HST
parallel fields, typically separated by 550 arcsec, are essentially
independent from each other. With independent fields the variance of
the total counts is given by the sum of the variances in each field
(as the set of counts in the fields are independent variables).

For higher redshift Lyman Break Galaxies, we plot in
Fig.~\ref{fig:zJdrop} the number counts uncertainty within both NICMOS
and WFC3 fields. A single NICMOS field has such a small area ($\approx
0.72$ arcmin$^2$) that $v_r$ is dominated by Poisson noise up to a few
counts per arcmin$^2$. The effect of cosmic variance in the number
counts uncertainty is instead more evident in a HST-WFC3 like field of
view, where the uncertainty is significantly higher than Poisson
noise. A very interesting conclusion that can be drawn from the figure
is that the number counts uncertainty in the six independent main and
parallel NICMOS fields, obtained through the HDF, HDF-South, UDF and
the UDF follow-up programs, is lower than that of one deep WFC3 field,
despite covering slightly less area. 
This example nicely shows that in order to minimize
cosmic variance effects in future surveys aimed at detecting $z>7$
Lyman Break Galaxies a sparse coverage is optimal. Of course a
continuous coverage has the advantage of enabling other science, such as
weak lensing studies at lower redshift.

\subsection{JWST surveys}

\begin{figure} 
  \plotone{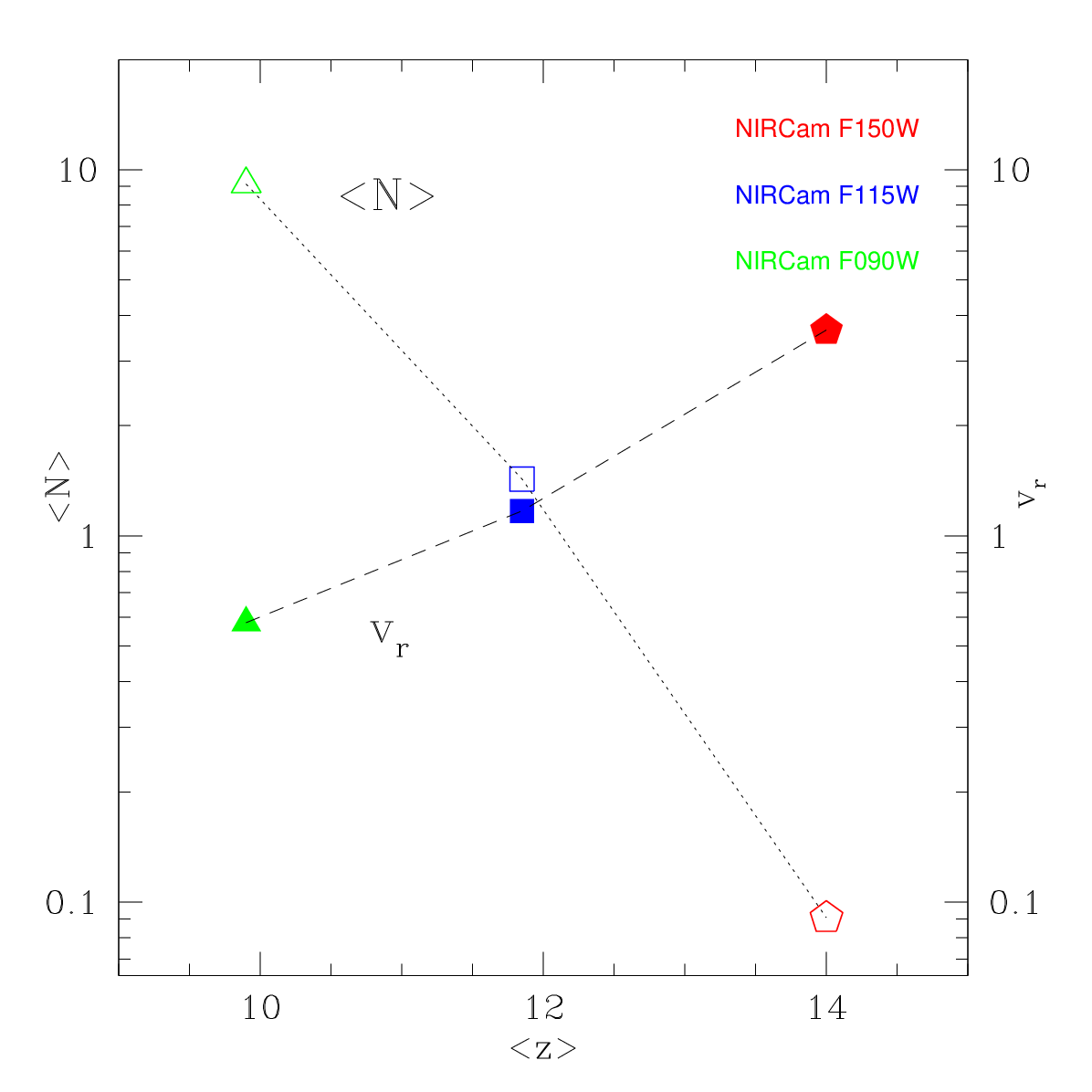} \caption{Total fractional error $v_r$ for JWST
  NIRCam dropouts (filled points) and average number of detections
  $\langle N \rangle$ per field of view (open points) versus the
  average redshift of the pass-band filter. The total field of view is
  given by two $2.2\times 2.2 arcmin^2$ fields separated by $30$
  arcsec. The counts go down to a mass limit equivalent to a depth of
  3 mag below $M_*$ for i-dropouts in a ACS field assuming the
  \citet{bou06} luminosity function. For F150W and F115W dropouts
  $v_r$ is dominanted by Poisson noise, while for F090W $v_r$ is about
  twice the Poisson error associated to the estimated 10 counts per
  field.}\label{fig:JWST}
\end{figure}

As a preliminary characterization of cosmic variance in future JWST
Lyman Break galaxy surveys, we present in Fig.~\ref{fig:JWST} the
cosmic variance for NIRCam F090W, F115W and F150W dropouts as given by
the combination of the two nearby field of views of NIRCam. This has
been obtained assuming no evolution from the $z=6$ $M/L$ relation
(eq.~\ref{eq:lm}). The light cone has been traced through snapshots in
our $680^3$ particles run, considering dark matter halos down to $50$
particles (that is to a mass limit of $2.2 \cdot 10^{10}
M_{\sun}/h$). If the same cutoff is applied to an ACS i-dropouts
survey, then we get about $250$ i-dropouts per field of view
(corresponding to a magnitude limit $m_{AB}=29.1$ using the
\citealt{bou06} luminosity function). As can be seen from
Fig.~\ref{fig:JWST} the rapid evolution of the dark matter halo mass
function greatly reduces the number of halos above the cut-off mass
within the pencil beam at $z \gtrsim 10$. Therefore the expected
number of F150W dropouts detections per NIRCam field is only $N
\approx 0.1$ and these observations would be affected primarily by
Poisson uncertainty. These numbers have been derived assuming a mass
luminosity relation of dropouts at $z>6$ consistent with that of
i-dropouts, as deep surveys with JWST are expected to reach about the
$z=6$ UDF sensitivity up to $z=20$. Of course, it is well possible
that the actual detections of F150W dropouts will be higher if these
primordial galaxies are more luminous than their $z\approx 6$
counterparts, similarly to what happens between $z=0$ and $z=6$ (e.g.,
see \citealt{cooray05}). In addition, for a precise estimate of the
expected number of these very high redshift objects in deep JWST
surveys a detailed modeling of the relation between intrinsic and
observed luminosity is required. At least dust reddening should
however play only a minor effect (see \citealt{tre06}).

\subsection{Ground based narrow band surveys}

If we consider narrow band searches for high redshift galaxies we
typically have a different beam geometry, given by a large field of
view with a small redshift depth. For example, \citet{ouc05} detect
more than 500 Lyman-$\alpha$ emitters at $z=5.7 \pm 0.05$ in a
$1~deg^2$ area. Under these conditions, we estimate $v_r \approx
0.18$, more than 3 times the Poisson uncertainty of the counts. The
large variance is given by a combination of a contiguous field of view
with a small redshift interval probed, which gives a cosmic volume
probed by this search roughly equivalent to that of a single $10
\times 16$ arcmin$^2$ i-dropouts GOODS field.

\section{Influence of Cosmic Scatter on Luminosity Function parameters}\label{sec:lf}

\subsection{Bright-faint counts and environment}\label{sec:LFsteep}

The main influence of the environment probed by a deep field is on the
observed number density of galaxies and therefore on the normalization
of the luminosity function. However, the shape of the luminosity
function can also be affected. This is the case in the local universe,
where it has been observed that the luminosity function in voids is
steeper than in the field (e.g., see \citealt{hoy05}), but this
possibility seems to be neglected when deriving the luminosity
function for Lyman Break galaxies (e.g., in \citealt{bou06}). 

\begin{figure} 
  \plottwo{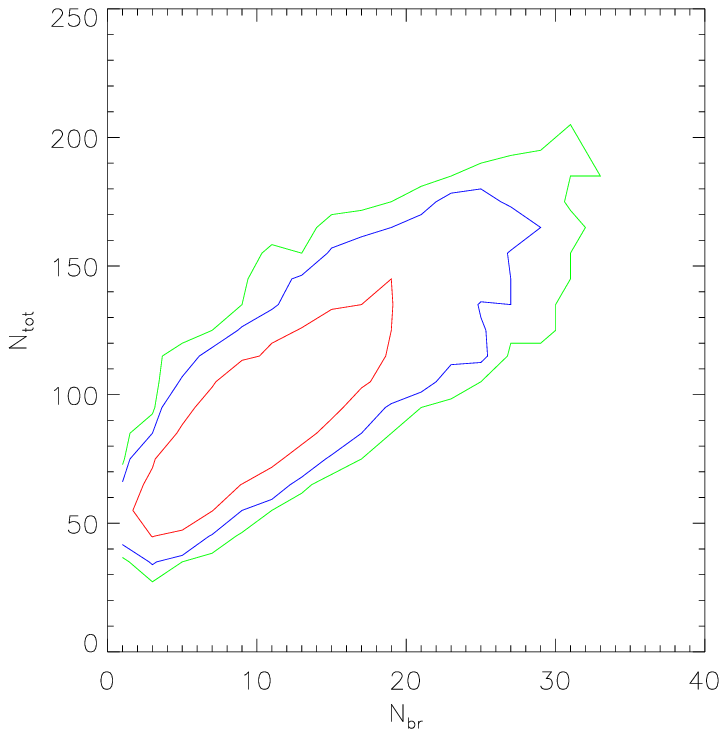}{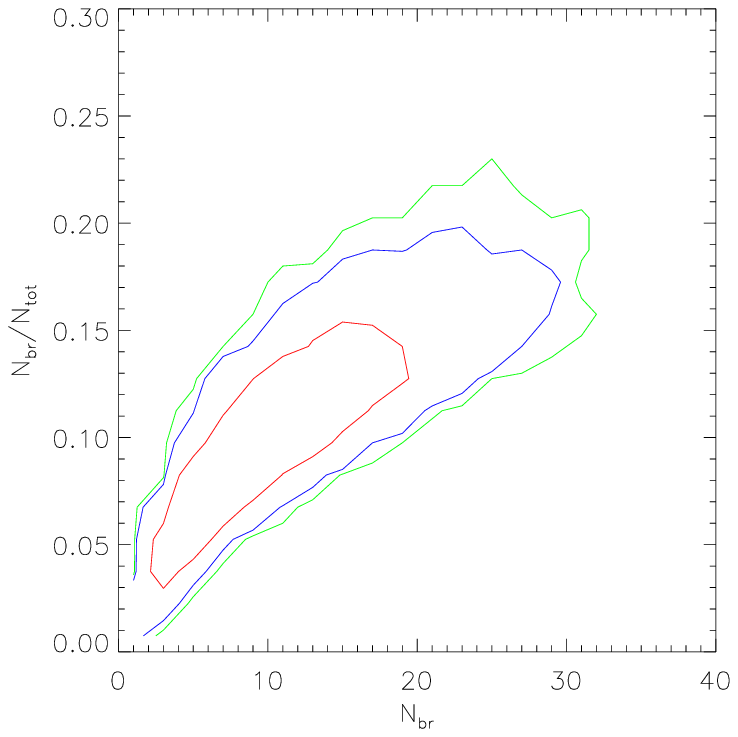}
  \caption{Distribution of the of the total ($N_{tot}$) versus bright
  ($N_{br}$) i-dropouts counts (left panel) and of the
  $N_{br}/N_{tot}$ ratio (right panel) simulated for one ACS field in
  function of the number of bright counts. The plots have been
  obtained using 4000 MC realizations. Confidence level contours are:
  red 68\%, blue 95\%, green 99\%. The average number of bright counts
  per field is $1.0 ~arcmin^{-2}$, while the average number of total
  counts is $8.6~arcmin^2$. From the right panel it is clear that the
  steepness of the luminosity function increases in under-dense
  regions.}\label{fig:slope}
\end{figure}

To highlight the importance of this effect even at high redshift, we
plot in Fig.~\ref{fig:slope} the distribution of massive halo
($M_{halo}>1.2 \cdot 10^{11} M_{\sun}$) counts versus total counts
down to a lower mass halo limit ($M_{halo}> 3 \cdot 10^{10} M_{\sun}
$) for a simulated i-dropouts survey with an ACS area. The average
number of total counts is roughly at the \citet{bou06} UDF depth (8.6
counts per $arcmin^2$), while the counts for the more massive halos
have an average density of $\approx 1 arcmin^{-2}$, which is
approximately the i-dropouts number density in GOODS. The ratio of the
massive to total counts can be adopted as an estimate for the
steepness (i.e. the shape) of the luminosity function. From the right
panel of Fig.~\ref{fig:slope} it is clear that when the beam passes
through under-dense regions the ratio easily decreases on average by
more than a factor two, although a large scatter is present. This
shows that the shape of the mass (luminosity) function does indeed
depend on the environment.

\begin{figure} 
  \plottwo{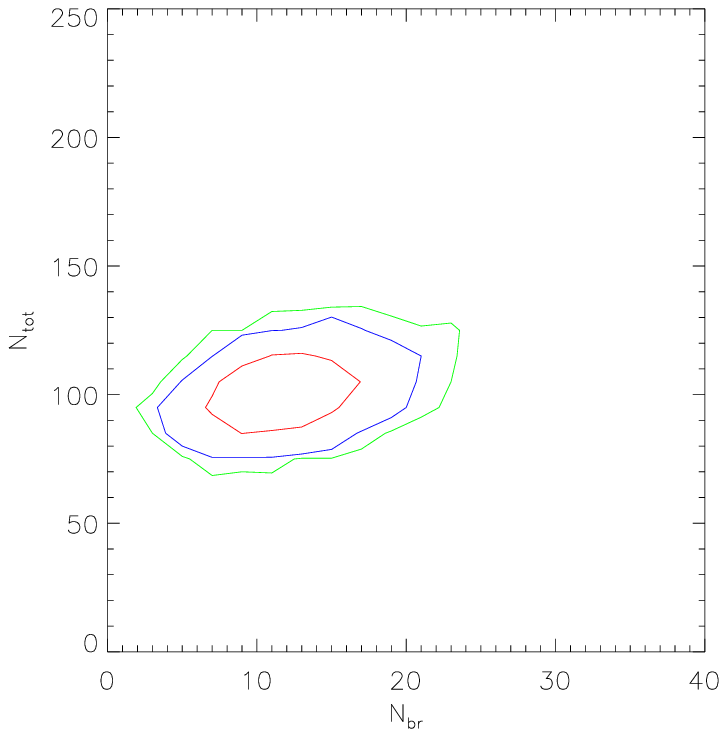}{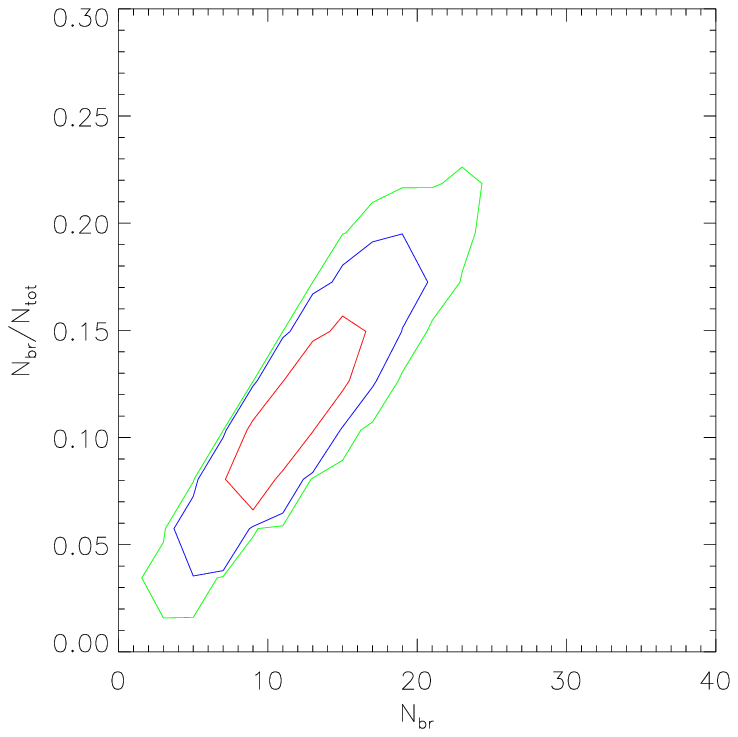}
  \caption{Like in Fig.~\ref{fig:slope} but for a toy model where
  $N_{ft}$ and $N_{br}$ are uncorrelated. Also here the luminosity
  function is steeper in underdense regions. }\label{fig:slope_poi}
\end{figure}

\begin{figure} 
  \plottwo{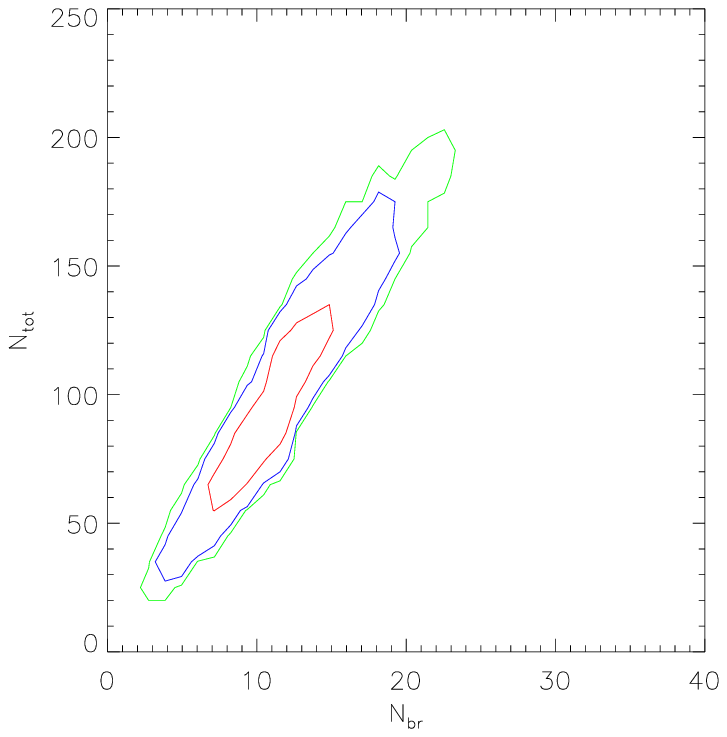}{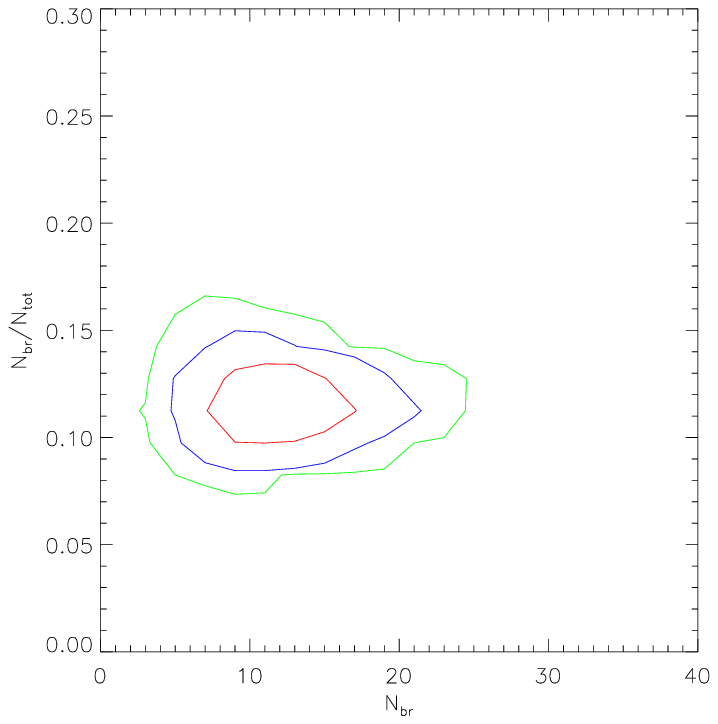}
  \caption{Like in Fig.~\ref{fig:slope} but for a toy model where
  $N_{ft}$ and $N_{br}$ are perfectly linearly correlated, apart from
  Poisson fluctuations in $N_{ft}$. This correlation leads to a
  luminosity function shape that is environment
  independent. }\label{fig:slope_corr}
\end{figure}

To understand better the origin of this effect we consider two
idealized scenarios to model the relation between bright and total
counts, where we consider only Poisson uncertainties. The first (see
Fig.~\ref{fig:slope_poi}) has been obtained by assuming that the
bright counts are completely uncorrelated from fainter ones. Therefore
we have:
\begin{equation}
\langle N_{tot} \rangle =  \langle N_{ft} \rangle +  \langle N_{br} \rangle,
\end{equation}
where $ \langle N_{ft} \rangle$ (that is the average difference
between the total and bright counts) depends on the minimum mass
cut-off mass considered. When $ \langle N_{ft} \rangle >> \langle
N_{br} \rangle $ bright and total counts are almost uncorrelated
between each other and therefore when a field is underdense in bright
counts, total counts are almost unaffected and its luminosity function
appears on average (much) steeper. The luminosity function is instead (quasi)
shape-invariant when we consider a model with a total correlation
between bright and total counts, where
\begin{equation}
\langle N_{tot} \rangle = \eta \langle N_{br} \rangle,
\end{equation}
with $\eta$ depending again on the minimum mass cut-off mass
considered. The results for this model are shown in
Fig.~\ref{fig:slope_corr} and have been obtained by first sampling the
bright counts ($N_{br~i}$) from a Poisson distribution with average
$\langle N_{br} \rangle$, and then sampling the faint counts
$N_{ft~i}$ from a Poisson distribution with average $p \times
N_{br~i}$ and $p=7.75$. This gives an average total number counts
$\langle N_{tot} \rangle = 8.75 \langle N_{br} \rangle $.

A realistic distribution of the counts, such as that obtained from our
mock catalogs is shown in Fig.~\ref{fig:slope} and lies between the
two extreme cases considered as toy models. In particular, the low
$N_{br}$ behavior of the $N_{br}/N_{tot}$ ratio is dominated by the
correlation with $N_{br}$, as can be seen by comparing
Fig.~\ref{fig:slope} with Fig.~\ref{fig:slope_poi}. This is because
the large scale structure introduces a correlation between bright and
faint counts but this correlation is not total. In fact deeper surveys
probe smaller mass halos, whose formation probability is sensitive to
higher frequencies in the power spectrum of primordial density
perturbations than that for more massive halos targeted in shallower
observations of the same field.

\subsection{Luminosity function shape and environment}

The results of the previous section suggest that a more thorough
characterization of the large scale structure influence on the mass
function at high redshift is required. Our main aim is to look for
systematic variations depending on the realized number counts value
and on the details of the fitting procedure employed. We fit a
Schechter function in the form:
\begin{equation} \label{eq:scheLF}
\phi(L) dL = \phi_* (L/L_*)^{\alpha} \exp([-L/L_*]) dL,
\end{equation}
to the distribution of galaxy luminosity derived from the dark matter
halo masses measured in our Monte Carlo code and transformed in
luminosities using the prescription of Eq.~\ref{eq:lm}. As discussed
in Sec.~\ref{sec:lm}, our treatment to build a sample of galaxy
luminosities is idealized and we are missing many observational
effects, such as apparent luminosity vs. redshift evolution and
redshift dependent selection effects within the redshift interval
considered. Our main aim is to highlight the importance of large scale
structure in the fitting of the luminosity function and not to
construct a detailed representation of observations.

\begin{figure} 
  \plotone{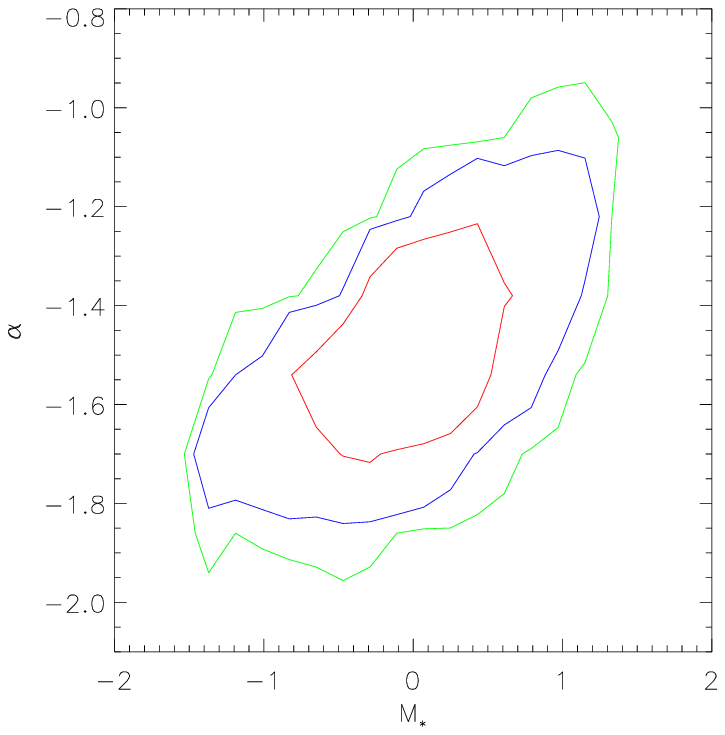} \caption{Best fitting Schechter
  function parameters for V-dropouts from one ACS field with 200
  average detections, simulated with 4000 realizations using our MC
  code. Parameters have been estimated using a Maximum Likelihood
  analysis.  Confidence level contours are: red 68\%, blue 95\%, green
  99\%. }\label{fig:binning}
\end{figure}

\begin{figure} 
  \plottwo{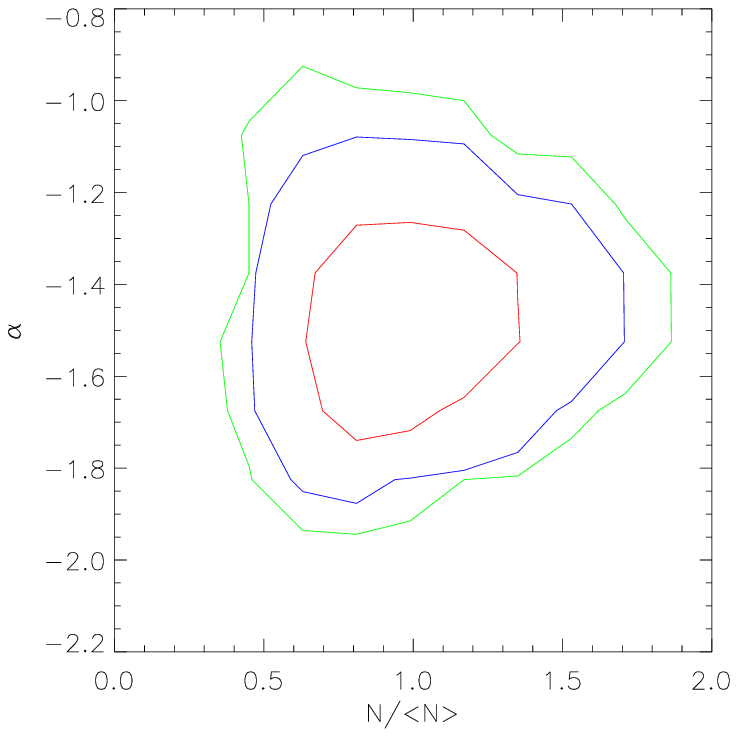}{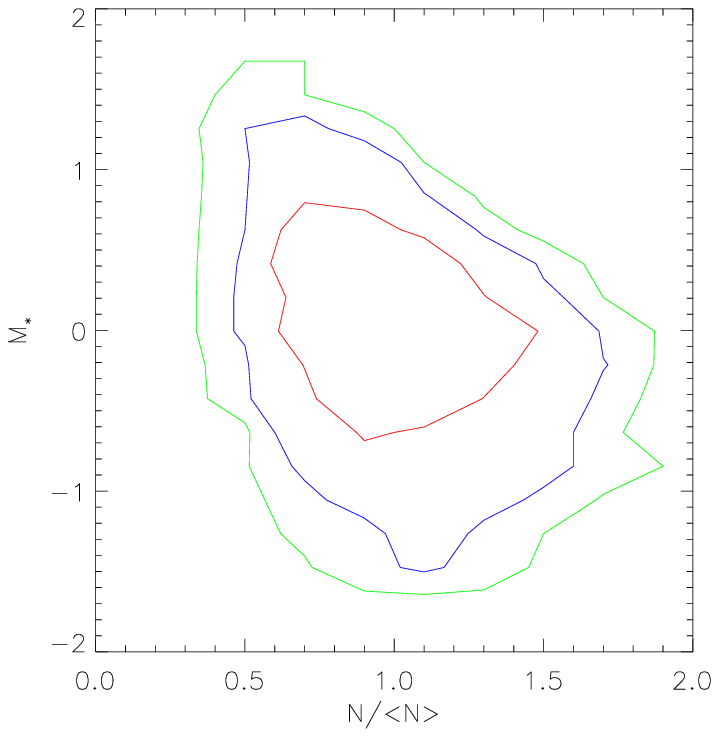}
  \caption{Maximum Likelihood best fitting Schechter function
  parameters as function of the number counts in the field of view for
  V-dropouts from one ACS field with an average of 200 detections,
  simulated with 4000 MC realizations. Left panel: $\alpha$; right
  panel: $M_{*}$.  Confidence level contours are: red 68\%, blue 95\%,
  green 99\%.}\label{fig:par_vs_N}
\end{figure}

We start by considering, for a simulated V-dropout deep sample in one
ACS field (200 galaxies on average in the pencil beam), the
distribution of the Schechter function parameter $L_*$ and $\alpha$
for 4000 Monte Carlo realizations\footnote{We recall here that not all
the 4000 realizations are truly independent as the total volume of the
box is only about 73 times larger than the pencil beam volume for
V-dropouts.}. We estimate the parameters using a standard Maximum
Likelihood estimator on the unbinned detections, following essentially
the procedure described in \citet{san79} (STJ79). For each synthetic
catalog $\{L_i\}_{i=1,N}$ we compute the likelihood for the luminosity
function in eq:~\ref{eq:scheLF}:
\begin{equation}
\mathfrak{L}(L_*,\alpha) = \prod_{i=1,N} \phi(L_i),
\end{equation}
where the $\phi_*$ is fixed by integrating the luminosity function up
to the detection limit $L_{min}$ of the survey and imposing the normalization:
\begin{equation}
\int_{L_{min}}^{+\infty} \phi(L) dL = 1.
\end{equation}
The maximization of the likelihood $\mathfrak{L}(L_*,\alpha)$ is then
carried out on a two dimensional grid with spacing $\Delta \alpha =
0.01$ and $\Delta M_* = 0.025$, where $M_*=-2.5~log_{10}{(L_*)}$. The
resulting distribution of best fitting parameters for the 4000
synthetic catalogs that we have generated are reported in
Fig.~\ref{fig:binning} and highlight the degeneration between $M_*$
and $\alpha$ parameters.
By plotting $L_*$ and $\alpha$ as a function of the number counts of the
survey (Fig.~\ref{fig:par_vs_N}) we can see that the slope of the
luminosity function does not depend on the environment, while $L_*$
does. When the number of counts in a field is above the average, $L_*$
is larger, although the scatter at fixed number of counts is
significant.

To further quantify the effect of cosmic scatter on the luminosity
function fitting, we consider the case where Lyman Break Galaxies are
detected both in a large area deep survey, such as GOODS, as well as
in a single, or few, pencil beams at a greater depth, such as the UDF
and UDF follow-up fields. This combination of data appears ideal from
the observational point of view as it allows us to constraint the
break $L_*$ of the luminosity function using detections from the large
area, shallower, survey, and the faint end slope using the deeper
dataset. However, the fitting procedure employed may lead to
artificial biases, especially when one tries to naively correct for
the effects of cosmic variance by considering its effect on $\phi_*$
alone. In fact, one might be tempted to re-normalize the luminosity
function for the deeper fields by considering the number of dropouts
galaxies detected at the same depth of the larger area, fainter
survey. For example, this is what has been done by \citet{bou06} (see
also \citealt{bou07}), who obtained the luminosity function for
i-dropouts by multiplying the UDF and UDF parallel fields counts by
the factors $1.3$ and $1.5$ (respectively) in order to account for a
deficit of bright detections with respect to the GOODS fields. This
correction has two potential problems that may contribute to
introducing an artificial steepening of the luminosity function and
both problems are apparent from Fig.~\ref{fig:slope}. First of all it
is clear from the left panel of Fig.~\ref{fig:slope} that if one were
to rescale the luminosity function of the deeper field, a relation of
the form $N_{ft} = \eta N_{br}$ would not be justified, as we show in
Sec.~\ref{sec:LFsteep}. In fact, the best fitting linear relation for
the counts in Fig.~\ref{fig:slope} is $N_{ft}=4.92 N_{br}+47$. This
implies that a 50\% deficit in bright counts with respect to the
average would correspond on average to only a $20\%$ deficit in faint
counts and not to the $50\%$ naive estimate of the deficit. There is
also a second, subtler effect introduced by a renormalization of the
luminosity function based on the number counts: $N_{br}/N_{ft}$ is
strongly correlated with $N_{br}$ in underdense fields due to the lack
of luminous galaxies, so a renormalization of the data based on
matching the $N_{br}$ counts to a reference value introduces an
artificial steepening of the faint end of the luminosity function.

As a quantitative example, we consider a test case where the
luminosity function is determined by GOODS-like data (one $10 \times
16$ arcmin$^2$ field) at the bright end and by one ACS field at the
faint end. The average number counts per $arcmin^2$ that we adopt are
$0.75$ for the bright end and $8.75$ for the deeper field (to be
consistent with the number densities in \citealt{bou06}). This gives
us an average of $120$ bright objects in the large area field and of
$100$ objects down to the fainter limit of the UDF-like field.

The luminosity function is then fitted by adopting three different
methods:
\begin{enumerate}
\item[(i)] first an observed luminosity function is constructed from
the synthetic catalogs using binned data (0.5 mag bins) by combining the
data from the two surveys with no Large Scale Structure correction,
as described in Sec 5.3 of \citet{bou06}. Then we fit the model
luminosity function using a maximum likelihood approach to these
binned data, that is we compute the theoretical expectation for the
number of objects in each magnitude bin and then we maximize the
likelihood under the assumption that the counts in each bin are
Poisson distributed.

\item[(ii)] maximum likelihood on binned data, as in point (i)
above, but here the luminosity function determination includes a large
scale correction a la Bouwens et al. (that is relying on the relation
$N_{ft} = \eta \cdot N_{br}$.), as in Sec. 5.1 of \citet{bou06}.

\item[(iii)] unbinned maximum likelihood modeling of the data with
free normalization of the data between the deep and wide fields. The
procedure is a straightforward generalization of what discussed above
for the determination of the luminosity function in a single field and
proceeds as described in \citet{san79}, where this method has been
first applied to luminosity function fitting\footnote{Here we would
like to stress that the \citet{san79} procedure relies on unbinned
data, so the STJ79 fitting adopted by \citet{bou07} is not really an
application of this method. Note also that Eq.~A6 in \citet{bou07} for
the probability distribution of the counts in each bin is not the
Poisson one and the maximization of their likelihood returns a result
equivalent to maximizing a Poisson probability distribution only in
the limit of a perfect data-model match, that is when the number of
observed objects in each bin is equal to the number of expected
objects. This difference is likely to affect the confidence regions
for the best fitting parameters.}.

\end{enumerate}

\begin{figure}
  \plottwo{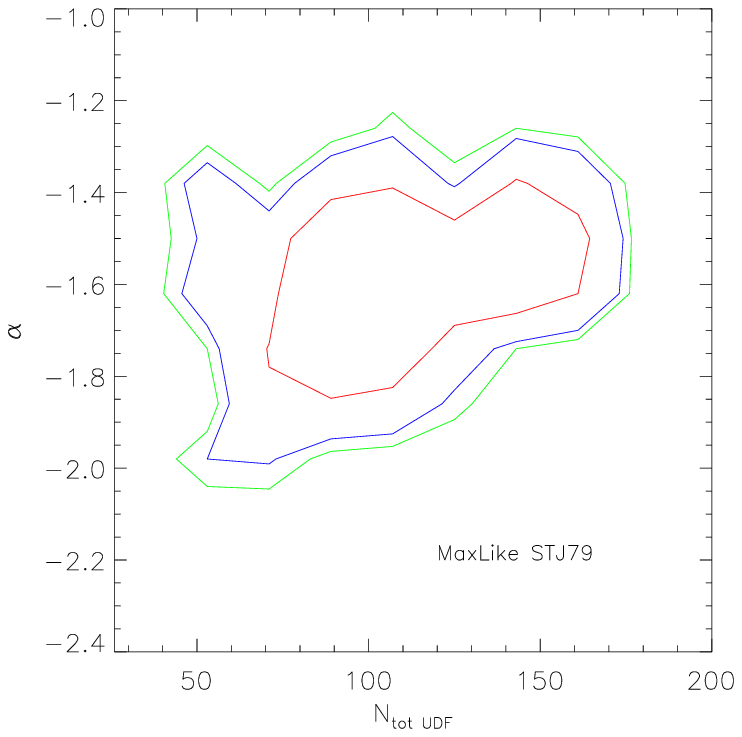}{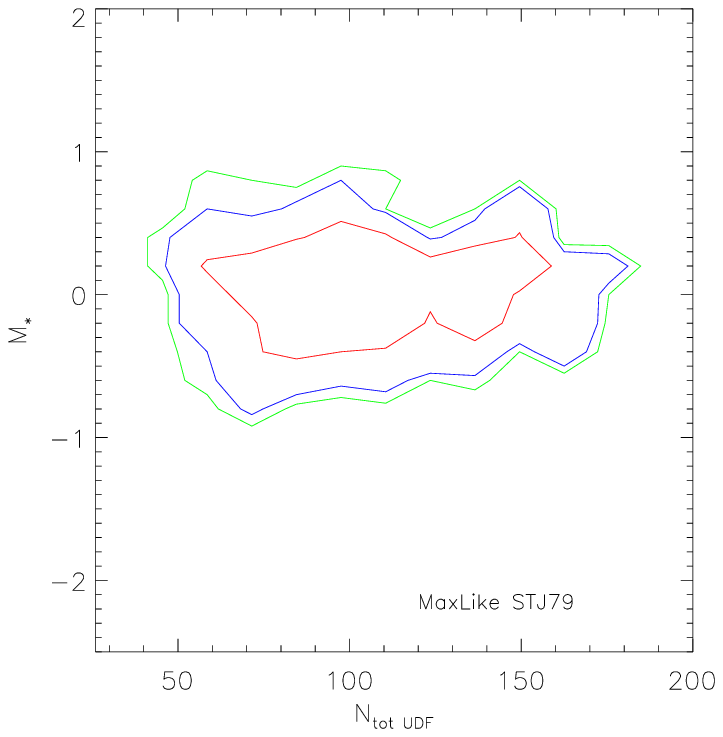}
  \plottwo{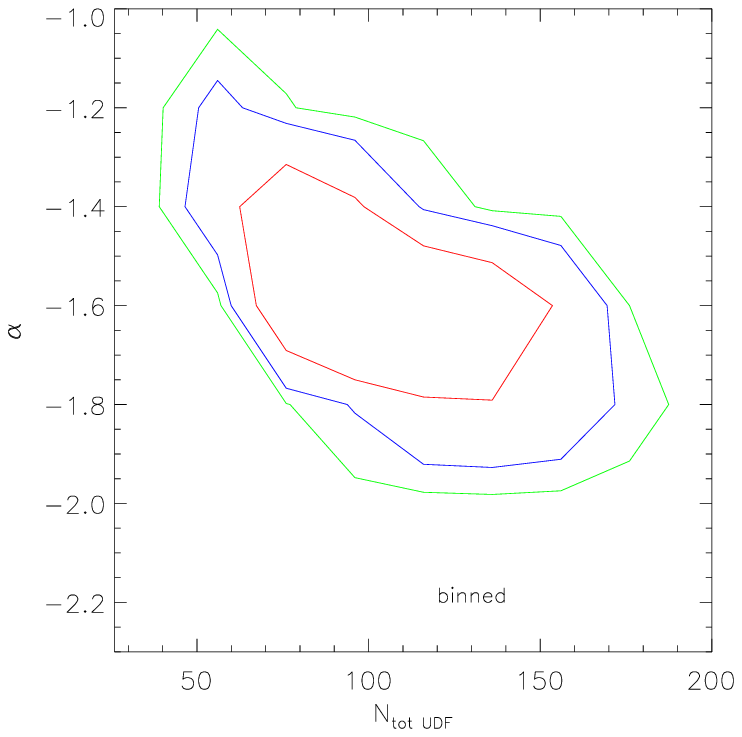}{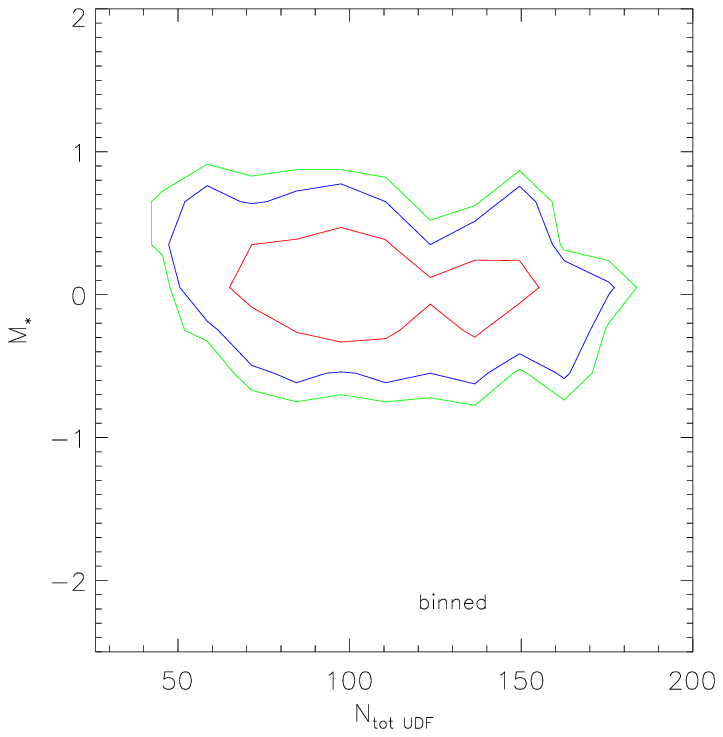}
  \plottwo{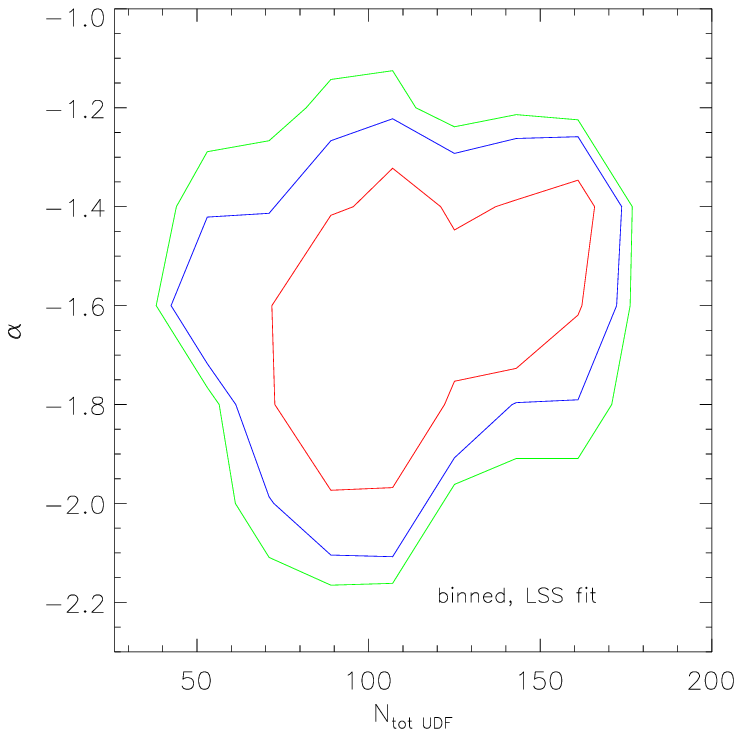}{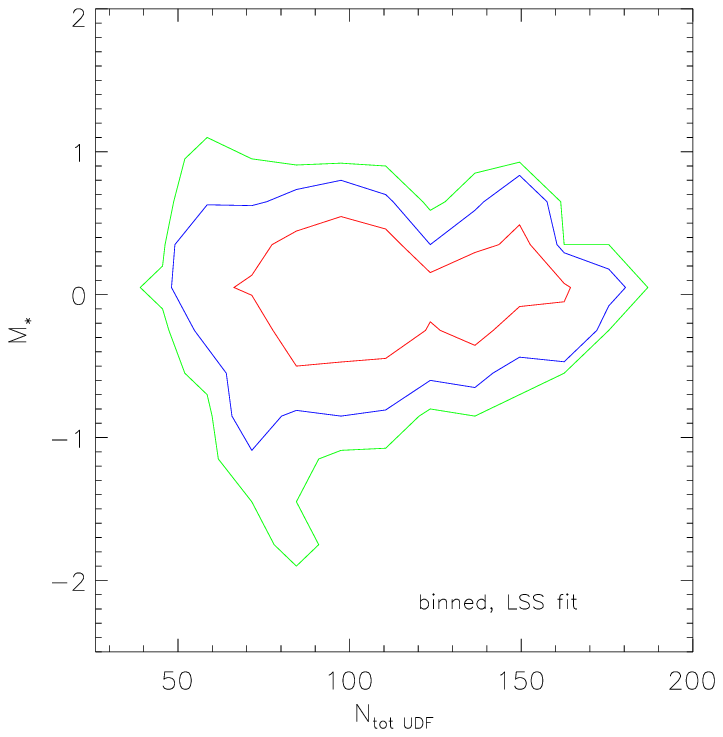}
   \caption{Best fitting parameters $\alpha$ (left) and $M_*$ (right)
  from unbinned i-dropout data (600 synthetic catalogs) that are a
  combination of one GOODS like field (with 120 objects on average)
  plus an UDF-like field (with 100 objects on average) for the three
  different fitting methods considered in the paper.}
  \label{fig:ML_idrop_combo}
\end{figure}

The results from these different fitting methods are shown in
Figs.~\ref{fig:ML_idrop_combo}, where we plot the contour levels of
the distribution of the best fitting luminosity function parameters
obtained from 600 different combinations of deep and wide fields. All
the parameters are shown as a function of the total number counts in
the deep field. The maximum likelihood method with free normalization
between the two fields \citep{san79} has a one sigma uncertainty of
$\approx 0.15$ on $\alpha$ and of $\approx 0.3$ on $M_*$. The
uncertainties are similar for when the fit is performed on binned data
without using the large scale renormalization of \citet{bou06}, but in
this case the fixed normalization between the two fields introduces a
bias in $\alpha$, with underdense deep fields leading to a shallower
slope. Applying the environment renormalization following
\citet{bou06} does over-correct the problem with a larger uncertainty
in $\alpha$ and $M_*$ ($0.21$ and $0.36$ respectively) and a
preference for steeper shapes of the luminosity function. This is an
artifact as the faint end of the luminosity function is overestimated
by the correction applied when the deep field is lacking luminous
galaxies. This behavior is also apparent from the fits performed in
\citet{bou07}:, e.g. in their Table 6 they estimate for i-dropouts
$\alpha = -1.77 \pm 0.16$ using a maximum likelihood approach similar
to \citet{san79} and $\alpha -2.06 \pm 0.20$ using the large scale
renormalization method.

\section{Conclusion}\label{sec:conc}

In this paper, we present detailed estimates of the variance in the
number counts of Lyman Break Galaxies for high redshift deep surveys
and of the resulting impact on the determination of the galaxy
luminosity function. The number counts distribution has been derived
from collisionless, dark matter only cosmological simulations of
structure formation assuming a $M/L$ relation and has been compared
with analytical estimates obtained from the two point correlation
function of dark matter halos. Halos have been identified in the
simulations snapshots, saved at high frequency ($\Delta z = 0.125$ up
to $z=10$) and a pencil beam tracer Monte Carlo code has been used to
construct synthetic catalogs of the halos above a minimum mass
threshold within the selection window for Lyman Break Galaxies at
different redshifts (corresponding to V,i,z and J dropouts for HST
surveys). In addition we also consider JWST NIRCam F090W, F115W and
F150W dropouts up to $z=15$.

By populating the dark matter halos with galaxies using different
semi-analytical prescriptions, that include multiple halo occupation
and different detection probabilities, we have shown that, to first
order, the standard deviation of the number counts for a given dropout
population depends mainly on the average value of number counts and on
the geometry of the pencil beam. This is because the average bias of
the population varies little as a function of the number density and is
reassuring for the robustness of our results, as it means that the
still uncertain details of high redshift galaxy formation are unlikely
to significantly affect the amount of cosmic variance in deep surveys.

The distribution of the number counts around its central value is
highly skewed for low number counts, while it becomes progressively
more symmetric as the average number of objects in the field of view
increases. The ratio of the measured variance to the variance expected
from Poisson noise is an increasing function of the average number of
objects in the field. For the typical number counts of V and i-dropout
in an ACS field of view at the UDF depth the one sigma fractional
uncertainty is about three times that due to Poisson noise. This has a
major impact on the rarity of overdensities of high redshift galaxies
such as those reported by \citet{mal05}, that while highly significant
with respect to a Poisson statistics, turn out to be not uncommon when
the effect of clustering is taken into account (e.g. at the $1.5
\sigma$ level in the \citealt{mal05} case).

The geometry of the volume probed is fundamental in defining the
number counts variance and a long and narrow beam has a lower variance
with respect to that estimated from an equivalent spherical (or
cubical) volume. This is because the long beam passes through many
different environments, while a spherical volume may happen to sit
right on top of extreme over-densities or under-densities. For example
for i-dropouts an ACS like field of view of $6 \times 6 \times 320
(Mpc/h)^3$ has a fractional uncertainty of $\approx 30\%$ for 100
counts on average, while an equivalent cubic volume would have an
uncertainty of $\approx 50 \%$.

Number counts in nearby fields are significantly correlated: for
i-dropouts, two adjacent ACS fields with 50 i-dropouts counts on
average, have a linear correlation coefficient $r_{lin}=0.41$. This
becomes $r_{lin} \lesssim 0.05$ at angular separation of about $550$
arcsec. This has two important consequences: (i) large area surveys
with adjacent exposures such as GOODS are still affected by a
non-negligible amount of cosmic variance, so that $v_r$ is of the
order of $\approx 20\%$ for i-dropouts down to its faint detection
limit and combining the two North and South fields, and (ii) HST
parallel observations, separated by about $600$ arcsec, such as those
from the UDF and UDF follow-up programs, can be considered essentially
independent fields. For independent fields the cosmic variance
decreases as the square root of the number of fields (in fact the
variance from independent variables sums up in quadrature). From the
point of view of future observations this also implies that the
currently existing six ultra deep NICMOS fields have a smaller total
cosmic variance with respect to a future single WFC3 deep field to
comparable depth, despite the greater area of the latter.

Using a simple mass-luminosity relation (Eq.~\ref{eq:lm}) we also
investigate the effects of cosmic variance on the determination of the
galaxy luminosity function. The impact of cosmic variance is not
limited on the normalization of the luminosity function, but extends
also to its shape. In fact, the luminosity function for under-dense
regions appears to be steeper than for field and cluster
environments. 
By fitting a Schechter function to
synthetic catalogs of V-dropouts galaxies in a UDF-like survey we find
that $M_*$ varies by about one magnitude from over-dense to
under-dense fields, while the slope $\alpha$ remains approximately
unchanged. An important caveat is that, as we have taken into account
essentially only dark matter clustering in our modeling, our result
could be changed if strong feedback effects due to baryon physics
are important.

This dependence of the luminosity function on the number counts of the
field has important consequences when attempts to correct for a
deficit of detections are made in dataset that combine a large area
survey with a small, deeper area. In fact, an artificial steepening of
the estimated luminosity function from binned data may arise when
naive corrections to account for under-densities are used, such as a
re-normalization of the faint end of the luminosity function in terms
of the ratio of bright counts in the deep area of the survey with
respect to the average value of bright counts over the whole survey
area. 

Therefore to determine the luminosity function for such survey
configurations the best approach appears to be the maximum likelihood
method applied to the unbinned data, as originally proposed by
\citet{san79}. The first step is to determine the shape ($\alpha$ and
$M_*$) of the Schechter function probability distribution, in case
convolved with a detection probability kernel to take into account
incompleteness and selection effects. This can be done by combining
the likelihood of both faint and bright counts, but allowing the
normalization $\phi_*$ to be a free parameter and to vary among the
two samples. Next, $\phi_*$ is estimated from the total number counts
of the two samples, compared with the expectation from integration of
the luminosity function between the relevant luminosity limits. This
method appears to be relatively unbiased with respect to cosmic
variance and represents an approach that does not require to quantify
the relative normalization of the detections in the different fields
considered.

Finally we stress that the intrinsic uncertainty due to cosmic
variance present while estimating the luminosity function parameters
must be taken into account when claims are made on the redshift
evolution of these parameters. For example, by combining one
GOODS-like field with a single deep, UDF like field, cosmic variance
introduces a $1\sigma$ uncertainty in $\alpha$ of $\Delta \alpha
\approx 0.15$. This error is a systematic contribution that comes on
top of any other contribution to the total error budget.

\acknowledgements

We thank Harry Ferguson for interesting and useful discussions and we
are grateful to the referee for a very careful reading of the
manuscript and for constructive suggestions. This work was supported
in part by NASA JWST IDS grant NAG5-12458, by STScI-DDRF award
D0001.82365 and by NCSA-Teragrid award AST060032T. A public version of
the cosmic variance calculator based on extended Press-Schechter
formalism is available at:
{\begin{verbatim} http://www.stsci.edu/~trenti/CosmicVariance.html \end{verbatim}}.
 

\clearpage
  
\clearpage

\clearpage
\clearpage

\clearpage

\clearpage


\clearpage

\clearpage

\clearpage

\clearpage

\clearpage

\clearpage

\clearpage

\clearpage

\clearpage

\begin{table}
\begin{center}
\caption{Summary of N-body simulations \label{tab:sim}}
\begin{tabular}{ccc}
\tableline\tableline
$N$ & $L_{box} [Mpc/h]$ & $\sigma_8$ \\
$512^3$  & $100$ & 0.75 \\
$512^3$ & $160$  & 0.75 \\
$680^3$  & $128$ & 0.9  \\

\tableline
\end{tabular}
\tablecomments{Summary of the N-body simulations of structure
formation. The first column reports the number of particles $N$ used,
the second the box edge size ($L_{box}$) and the third the
normalization $\sigma_8$ of the amplitude of the power spectrum of
density fluctuations.}
\end{center}
\end{table}


\begin{table}
\begin{center}
\caption{Pencil Beam Properties \label{tab:fields}}
\begin{tabular}{lccccc}
\tableline\tableline
~~~~~ID & Angular Size [('')$^2$] & $z_{min}$ & $z_{max}$ & Comoving Size [(Mpc/h)$^3$] & Sep. \\
ACS v-drop & $205 \times 205$ & $4.6$ & $5.7$ & $5.7 \times 5.7 \times 420$& -\\ 
ACS i-drop &  $205 \times 205$ & $5.65$ & $6.7$ & $6.0 \times 6.0 \times 320$& -\\ 
GOODS i-drop &  $600 \times 960$ & $5.65$ & $6.7$ & $17.6 \times 28.1 \times 320$& -\\ 
NIC3 z-drop & $51 \times 51$ & $7.0$ & $8.5$ & $1.6 \times 1.6 \times 342$& -\\ 
WFC3 z-drop & $125 \times 137 $ & $7.0$ & $8.5$ & $3.9 \times 4.3 \times 342$& -\\ 
NIC3 J-drop & $51 \times 51$ & $8.5$ & $10.0$ & $1.7 \times 1.7 \times 269$& -\\ 
WFC3 J-drop & $125 \times 137 $ & $8.5$ & $10.0$ & $4.1 \times 4.5 \times 269$&-\\ 
\hline
JW\_F090W-drop & $2 \times(130 \times 130) $ & $9.0$ & $10.8$ & $2 \times (4.5 \times 4.5 \times 295)$ & 30'' $\equiv$ 1.0 Mpc/h\\ 
JW\_F115W-drop & $2 \times(130 \times 130) $ & $11.0$ & $12.7$ & $2 \times (4.6 \times 4.6 \times 217)$& 30'' $\equiv$ 1.0 Mpc/h\\ 
JW\_F090W-drop & $2 \times(130 \times 130) $ & $12.9$ & $15.1$ & $2 \times (4.8 \times 4.8 \times 223)$& 30'' $\equiv$ 1.1 Mpc/h\\ 

\tableline
\end{tabular}
\tablecomments{Assumed properties for the Pencil Beam Fields used
throughout this paper.}
\end{center}
\end{table}


\begin{table}
\begin{center}
\caption{Cosmic variance for the models in Fig.~\ref{fig:idrop_models} \label{tab:uncert}}
\begin{tabular}{lc}
\tableline\tableline
~~~~~ID & $v_r$ \\
$f_{ON}=0.25$, HighRes & 0.32\\
$f_{ON}=0.5$ & 0.35\\
$f_{ON}=1$ &  0.39 \\
$f_{ON}=1$, HOD & 0.42 \\
$f_{ON}=1$, HighRes & 0.40 \\
$f_{ON}=1$, HighRes, MagCut & 0.42 \\

\tableline
\end{tabular}
\tablecomments{Cosmic variance $v_r$ for i-dropouts (40 counts on
average) in one ACS field for the different models presented in
Fig.~\ref{fig:idrop_models}.}
\end{center}
\end{table}


\end{document}